\documentclass{aa}
\usepackage{graphicx}
\usepackage{txfonts}
\usepackage{hyperref} 

\usepackage{xspace}
\usepackage{subcaption}

\def\grb{$\gamma$-ray binary\xspace}
\def\ltsima{$\; \buildrel < \over \sim \;$}
\def\lsim{\lower.5ex\hbox{\ltsima}}
\def\gtsima{$\; \buildrel > \over \sim \;$}
\def\gsim{\lower.5ex\hbox{\gtsima}}

\def\lsi{LS~I~+61$^\circ$~303\xspace}
\def\ls{LS~5039\xspace}
\def\psrb{PSR~B1259-63\xspace}
\def\fgl{1FGL~J1018.6-5856\xspace}
\def\etacar{$\eta$~Carinae\xspace}
\def\lmc{LMC P3\xspace}
\def\gamvel{$\gamma^2$~Vel\xspace}

\def\hess#1#2#3#4{\ifnum#1#2#3#4=0632
          HESS~J0632+057\xspace
       \else
          \ifnum#1#2#3#4=1832
            HESS~J1832-093\xspace
          \else
            HESS~J#1#2#3#4(??)\xspace
       	   \fi
           \fi}

\def\fermi{\textit{Fermi}\xspace}
\def\flat{\textit{Fermi}-LAT\xspace}
\def\swift{\textit{Swift}\xspace}
\def\xmm{\textit{XMM-Newton}\xspace}
\def\cha{\textit{Chandra}\xspace}
\def\nus{\textit{NuSTAR}\xspace}

\hypersetup{draft}   
\begin{document}

   \title{Overview of non-transient $\gamma$-ray binaries and prospects for the Cherenkov Telescope Array}
   \authorrunning{M. Chernyakova et al.}


   \author{M. Chernyakova
          \inst{1,2}, 
          D. Malyshev\inst{3},
          A. Paizis\inst{4}, N. La Palombara\inst{4}, M. Balbo\inst{9}, R. Walter\inst{9}, B. Hnatyk\inst{10}, B. van Soelen\inst{11}, P. Romano\inst{5},
          P. Munar-Adrover\inst{6}, Ie. Vovk\inst{7},  G. Piano\inst{8}, F. Capitanio\inst{8},
        D.~Falceta-Gon\c calves\inst{16}, M.~Landoni\inst{5}, P.~L.~Luque-Escamilla\inst{15}, J. Mart\'i\inst{15}, J. M. Paredes\inst{12}, M. Ribó\inst{12}, S. Safi-Harb\inst{13}, L. Saha\inst{14},  L. Sidoli\inst{4}, S.~Vercellone\inst{5}
          }

   \institute{School of Physical Sciences and CfAR, Dublin City University, Dublin 9, Ireland\\
              \email{masha.chernyakova@dcu.ie}
         \and
         Dublin Institute for Advanced Studies, 31 Fitzwilliam Place, Dublin 2, Ireland
         \and 
             Institut f{\"u}r Astronomie und Astrophysik T{\"u}bingen, Universit{\"a}t T{\"u}bingen, Sand 1, D-72076 T{\"u}bingen, Germany 
         \and 
         INAF - IASF Milano, via Alfonso Corti 12, I-20133 Milano, Italy
         \and 
          INAF - Osservatorio Astronomico di Brera, 
          Via E.\ Bianchi 46, I-23807 Merate, Italy   
         \and
         Unitat de F\'isica de les Radiacions, Departament de F\'isica, and CERES-IEEC, Universitat Aut\`onoma de Barcelona, E-08193 Bellaterra, Spain
         \and 
         Max-Planck-Institut f\"ur Physik, D-80805 M\"unchen, Germany
         \and 
         INAF – IAPS Roma
         \and 
         ISDC, University of Geneva
         \and 
         Astronomical Observatory of Taras Shevchenko National University of Kyiv,
3 Observatorna str. Kyiv, 04053, Ukraine
          \and
           University of the Free State Department of Physics PO Box 339 9300 Bloemfontein South Africa. 
           \and 
         Institut de Ciències del Cosmos (ICCUB), Universitat de Barcelona, IEEC-UB, Martí i Franquès 1, E-08028 Barcelona, Spain
           \and 
           Dept of Physics and Astronomy, University of Manitoba, Winnipeg, Canada, R3T 2N2
           \and
           Unidad de Partículas y Cosmología (UPARCOS), Universidad
Complutense, E-28040 Madrid, Spain
           \and
           EPS Ja\'en, Universidad de Ja\'en, Campus Las Lagunillas s/n 23071 Ja\'en, Spain
           \and
           Escola de Artes, Ci\^ encias e Humanidades, Universidade de S\~{a}o Paulo, Rua Arlindo Bettio 1000, CEP 03828-000 S\~{a}o Paulo, Brazil
}


 
    \abstract
    {}
    {
        Despite recent progress in the field, there are still many open questions regarding $\gamma$-ray binaries.
        In this paper we provide an overview of  non-transient $\gamma$-ray binaries and discuss how 
        observations with the 
        Cherenkov Telescope Array (CTA) will contribute to their study.
    }
    {
        We simulate the spectral behaviour of the non-transient $\gamma$-ray binaries  
        using  archival observations as a reference. 
        With this we test the CTA capability to measure the sources' spectral parameters and detect variability on various time scales.
    }
    {
        We review the known properties of $\gamma$-ray binaries and the theoretical models that have been used to describe their spectral and timing characteristics. We show that CTA is capable of studying these sources on  time scales comparable to their characteristic variability time scales. 
        For most of the binaries, the unprecedented sensitivity of CTA will allow the spectral evolution to be studied on a time scale as short as 30 min. This will enable a direct comparison of the TeV and lower energy (radio to GeV) properties of these sources from simultaneous observations. We also review the source-specific questions that can be addressed with such high-accuracy CTA measurements.
    }
    {}


   \maketitle
%
%

\section{Introduction}

Gamma-ray binaries are a subclass of high-mass binary systems whose energy spectrum peaks at  high energies (HE, E$\gtrsim$100 MeV) and extends up to very high energy (VHE, E$\gtrsim$100 GeV) $\gamma$ rays. In these systems a compact object (neutron star, NS, or a black hole, BH) is orbiting around a young massive,  either O- or B- type, star.

While high-mass binaries represent a substantial fraction  of Galactic X-ray sources detected above 2 keV  \citep[e.g.][]{Grimm2002}, less than ten binaries were detected in the $\gamma$-ray band  by the current generation of Cherenkov  telescopes, such as MAGIC \citep{2012APh....35..435A}, VERITAS \citep{2015ICRC...34..771P} and  H.E.S.S. \citep{2006A&A...457..899A}. As such,  $\gamma$-ray binaries represent a relatively new  {and unexplored} class of astrophysical objects.

Among all the binary systems regularly observed at TeV energies, the nature of the compact object is only firmly established for two systems, namely \psrb and PSR J2032+4127. In \psrb, a 43 ms radio pulsar is orbiting around a Be star, in a very eccentric 3.4 years orbit which, except for a brief period near periastron, allows radio pulsations to be detected.   
The second source, which also contains a pulsar, PSR J2032+4127, has an even longer orbital period of about 50 years \citep{PSRJ2032_Lyne_2015,PSRJ2032_Ho_2017} and TeV emission was detected from this system by VERITAS and MAGIC as the pulsar approached its periastron in September 2017 \citep{PSRJ2032_VHE_2017ATel}. 

All other known systems are more compact and the nature of their compact object is yet unknown. It is possible that these systems harbour radio pulsars as well, but the optical depth due to the stellar 
wind outflow is too high to detect the radio signal originating close to the pulsar (``hidden pulsar model'', see e.g. \citealt{Zdziarski_2010MNRAS_LSI}). Alternatively, it is possible that some of these systems harbour a black hole or an accreting neutron star (``microquasar model'', \citealt{1998Natur.392..673M}).

Among the accretion powered $\gamma$-ray binaries which likely  {contain} a black hole, the highest energy emission which has so far been regularly observed is 
from Cyg X-1 and Cyg X-3, detected with AGILE and Fermi Large Area Telescope (\textit{Fermi}-LAT) \citep[e.g.][]{2009Natur.462..620T,2010ATel.2715....1S,2013ApJ...775...98B, 2013MNRAS.434.2380M}. Cyg X-1 is detected up to about 10~GeV in the hard state \citep{CygX1-2016-Zanin,CygX1_2017_Zdz}, and Cyg X-3 is detected up to about 10 GeV during flares that mostly occur when the source is in the soft state \citep{2018MNRAS.479.4399Z}.  In 2006 MAGIC telescope has also reported marginal detection of TeV flare from Cyg X-1  at a 3.2 $\sigma$ confidence level  coinciding with an X-ray flare seen by RXTE, Swift, and INTEGRAL \citep{2007ApJ...665L..51A}.   Thus, at the moment, we have 
evidence that accreting sources can accelerate particles only during some very specific states, and we need to study the persistent  $\gamma$-ray binaries in detail to unveil their ability to steadily accelerate particles (at least at given orbital phases). 
Recently 
the microquasar SS433 was also detected
by \flat and HAWC \citep{2015ApJ...807L...8B,2018Natur.562...82A}. The persistent emission reported by HAWC is localised to structures in the lobes, far from the centre of the system. This implies  an emission scenario  very different to the rest of the systems.

In addition to the   $\gamma$-ray binaries which contain a compact object, HE and VHE  $\gamma$-rays have been also detected from colliding wind binaries (CWB). A colliding wind binary is a binary star system consisting of two non-compact massive stars which emit powerful stellar winds, with 
large mass-loss rates and high wind velocities. The collision of the winds produces two strong shock fronts, one for each wind, both surrounding a shock region of compressed and heated plasma, where particles are accelerated to very high energies \citep{eichler93}.  Please note that 
to date only one CWB 
has a confirmed detection at VHE, namely  $\eta$ Carinae ($\eta$ Car). 

In the coming years the family of $\gamma$-ray binaries can be extended via follow-up observations based on indications from lower energy bands. In particular, binaries with pulsars orbiting Be or O-stars are likely to provide a noticeable addition to the $\gamma$-ray binary list. Similarly, black hole/Be-star binaries can also be considered as serious candidates to be admitted in the $\gamma$-ray binary family \citep{2010ApJ...723L..93W, 2016ApJ...829..101M}, although only one system, MWC 656, is known so far \citep{CasaresMWC6562014}. Other alternative search efforts have focused on multi-wavelength cross-identification that explores the possible association of luminous early-type stars with GeV $\gamma$-ray sources (mainly) detected by  \citep{2013arXiv1303.2018M, 2017A&A...598A..81M}. 
Still, recently \citet{2017A&A...608A..59D} carried out a synthetic population simulation, estimating that less than 230 systems
exist inside the Milky Way.

In recent years  
$\gamma$-ray binaries have already been the subject of numerous 
observational campaigns
and theoretical studies
\citep[e.g.][]{2013A&ARv..21...64D},
which strongly indicate 
that the high-energy emission from these systems is primarily powered by the outflow from the compact object. However, 
due to the limited sensitivity of the current generation of instruments the nature of the compact object (NS or BH), and the  details of the particle acceleration, with efficiency sometimes close to  the theoretical limit \citep[e.g][]{2018ApJ...863...27J},  remain unknown in the majority of the systems \citep[e.g][]{paredes19}.

Existing data already show that in some systems, like \ls and \lsi, the observed HE and VHE emission are separate components generated at different places \citep[e.g.][and references therein]{2013A&A...551A..17Z}. A proper modelling of such  double-component spectra requires time-resolved spectroscopy along the binary orbit.  In addition $\gamma$-ray binaries are known  to be variable on 
time scales as short as   {hours, minutes and even tens of seconds, as observed in X-ray and HE bands \cite[e.g.][]{2009MNRAS.397.2123C, 2009ApJ...693.1621S, 2018ApJ...863...27J}}. At the same time 
with the current generation of VHE telescopes 
observations can only provide information averaged over several days even for the brightest binaries. The possibility to study the broad band spectral variability on a characteristic time scales is crucial for  an unambiguous modelling.  

In the next decade this situation may change with the deployment of the next-generation 
VHE telescope, the Cherenkov Telescope Array (CTA) observatory. 
CTA will be composed of two sites, one in the Northern (La Palma, Canary Islands) and one in the Southern hemispheres (Paranal Observatory, Chile), which will enable observations to cover the entire Galactic plane and a large fraction of the extra-galactic sky \citep[see e.g.][]{CTA_2017}.  The array will include three different telescope sizes to maximise the energy range of the instrument (from 20 GeV to more than 300 TeV). With more than 100 telescopes in the Northern and Southern hemispheres, CTA will be the largest ground-based  $\gamma$-ray observatory in the world. CTA will be a factor of five to twenty times more sensitive (depending on the energy) than the current generation of
ground-based  $\gamma$-ray detectors \citep{2017arXiv170907997C}. 

It is foreseen that CTA will make a breakthrough in many areas, including the study of  $\gamma$-ray binaries. Beyond  detailed studies of the known binaries, CTA is foreseen to discover new sources, enlarging the population.   {\citet{2017A&A...608A..59D} has estimated that  four new gamma-ray binaries can be expected in the first two years of the CTA Galactic Plane survey.}

The aim of this paper is to estimate CTA's potential for the observations of known $\gamma$-ray binary systems. The text is organised as follows. In Sect.~\ref{sec::simulations} we outline the source selection and CTA simulation setup. Sections~\ref{sec::grlb} and~\ref{sec::cwb} present the results of simulation for specific binary system types. Finally, in Sect.~\ref{sec::summary} we briefly summarise and discuss the obtained results.

\begin{table*}
\caption{Properties of  $\gamma$-ray binaries with a compact source \label{parameters}}
\small
\begin{tabular}{llllllll}
\hline 
 & PSR & LS & LS I & HESS & 1 FGL & HESS& LMC P3 $^{**}$\\
 & B1259-63$^{\star}$ & 5039$^{\dagger}$ & 61$^\circ$ 303$^{\bullet}$ & J0632+057$^{\diamond}$ & J1018.6-5856$^{\ddagger}$ & J1832-093 &\\
 \hline
 P$_{\rm orb}$ (days) & 1236.724526(6)	&  3.90603(8) 		&  26.496(3) 	& 315(5) 		& 16.544(8) &-&10.301(2)\\
 $e$ 				& 0.86987970(6) 	&  0.24(8) 		& 0.54(3) 		& 0.83(8) 		& 0.31(16) &-& 0.40(7)\\
 $\omega$ ($\degr$)	& 138.665013(11) $^\sharp$ 		&  212(5) 		& 41(6) 		& 129(17)  & 89(30) &-& 11(12)\\
 $i$ (\degr)		& 153.3$^{+3.2}_{-3.0}$			& 13--64 			& 10--60 		& 47--80 		& -&-&-\\
 $d^1$ (kpc)		& 2.39$\pm{0.18}$ 			    & 2.07 $\pm$ 0.22  	& 2.63$\pm$0.26 & 2.76 $\pm$ 0.34 &6.52$\pm$1.08&-&50.0 $\pm$ 1\\
 \hline
spectral type 			& O9.5Ve 			& O6.5V(f)		& B0Ve 		& B0Vpe 		& O6V(f) &-& O5 III(f)\\
 $M_{\star}$ (M$_{\odot}$)  	& 14.2-29.8	& 23 			& 12 		& 16 		& 31 &-& -\\
$R_{\star}$ (R$_{\odot}$) 	& 9.2 	& 9.3 			& 10 		& 8 			& 10.1 &-&-\\
$T_{\star}$ (K) 				&  33500 & 39000 			& 22500 		& 30000 		& 38900 &-&40000\\
 \hline
$d_{\rm periastron}$ (AU) 	& 0.94		& 0.09  	& 0.19 & 0.40 & (0.41)&-&-\\
$d_{\rm apastron}$ (AU) 		& 13.4		& 0.19 	& 0.64 & 4.35 & (0.41)&-&-\\
$\phi_{\rm periastron}$ 		& 0		& 0		& 0.23	& 0.967 & - &-& 0.13\\
$\phi_{\rm sup.\ conj.}$ 		& 0.995	& 0.080	& 0.036	& 0.063 & -&-& 0.98\\
$\phi_{\rm inf.\ conj.}$ 		& 0.048	& 0.769	& 0.267	& 0.961 & -&-&0.24\\
 \hline
 IRF: 	  &South\_z40 & South\_z20& North\_z20& South\_z40&South\_z40 & South\_z20&South\_z40\\
          & & North\_z40& & North\_z20&  &North\_z40 &  \\
 \hline
\multicolumn{6}{l}{ $\star$ \citet{2014MNRAS.437.3255S,PSRB1259-2018_distance,2011ApJ...732L..11N}}\\
\multicolumn{6}{l}{ $\dagger$ \citet{2002A&A...384..954R,2004ApJ...600..927M,2011MNRAS.411.1293S}}\\ 
\multicolumn{6}{l}{ $\bullet$ \citet{2004ApJ...600..927M,2009ApJ...698..514A}}\\
\multicolumn{6}{l}{$\diamond$ 
\citet{casares12,Aliu2014HESS0632,2010ApJ...724..306A}}\\
\multicolumn{6}{l}{$\ddagger$ \citet{An+15,Monageng+17,Napoli+11}}\\
\multicolumn{6}{l}{$^{**}$ \citet{2016ApJ...829..105C,2013Natur.495...76P}}\\
\multicolumn{6}{l}{$\sharp$ argument of periastron of the pulsar orbit (massive star for the others systems)}\\
\multicolumn{6}{l}{$^1$All distances given with an error are taken from the Gaia archive, https://gea.esac.esa.int/archive/}
\end{tabular}
\end{table*}

\section{Simulations}
\label{sec::simulations}

All the simulations that are reported in this paper were performed with the {\tt ctools} analysis package\footnote{\href{http://cta.irap.omp.eu/ctools/}{http://cta.irap.omp.eu/ctools/}} \citep[][ v 1.5]{2016A&A...593A...1K}, 
together with the prod3b-v1 set of instrument response functions (IRFs\footnote{\href{https://www.cta-observatory.org/science/cta-performance/}{https://www.cta-observatory.org/science/cta-performance/}}) 
for both the Northern (La Palma) and Southern (Paranal) CTA sites. 
Note that in  prod3b-v1 IRFs only exist for zenith angles of 20 and 40 degrees. To select the correct response function we used a simple relation between the minimal source's zenith angle (mza) and declination (dec), and  latitude (lat) of the site: $mza=|lat - dec|$. For example, for  La Palma  
(lat = $+29^\circ$), HESS J0632+057 has a minimal zenith angle of 23 degrees. Thus the 20 degree IRF is the most appropriate.
For the Southern site (Paranal, 
lat = $-25^\circ$), HESS J0632+057 has a minimal zenith angle of 31 degrees, and we choose the 40 degree IRF. 

In the analysis,  we simulated the data with \textit{ctobssim} and fitted simulated event files  with \textit{ctlike} using a maximum likelihood method. To simulate the event file, we have used all sources 
listed in TeVCat \footnote{\href{tevcat2.uchicago.edu}{http://tevcat2.uchicago.edu/}} within a circle of $5^\circ$ around the source position. 
 In addition, we have included the instrumental background and the Galactic diffuse  $\gamma$-ray emission in the model of the region surrounding the simulated source.\footnote{We have verified that results obtained have a negligible dependency on the choice of the  Galactic diffuse  background model.}  
 Please note that all error presented in the paper are the statistical errors at a $1~\sigma$ confidence level.


\section{$\gamma$-ray binaries with a compact source. \label{sec::grlb}}

This section is devoted to an overview of the non-transient, point-like, $\gamma$-ray binary sources that all consist of an O or B/Be type star and a compact object (pulsar or black hole). 
The specific sources studied here are listed in Table \ref{parameters}.  PSR J2032+4127 is not included since, with a $\approx 50$ year orbital period, it is unlikely that the next periastron passage will be observed with CTA.  {Very recently, while this paper was under revision, a new gamma-ray binary candidate, 4FGL J1405.1-6119 was discovered \citep{2019arXiv190810764C}. At the moment TeV properties of the source are  not known and thus it is not discussed in this paper either.}

The VHE emission of all the $\gamma$-ray binaries is well described by a power-law  with an exponential high-energy cut-off. As was mentioned in the introduction current VHE observations are not sensitive enough to follow the details of the spectral evolution of these systems on their characteristic time scales.

\begin{figure*}
  \centering
  \includegraphics[width=0.95\columnwidth]{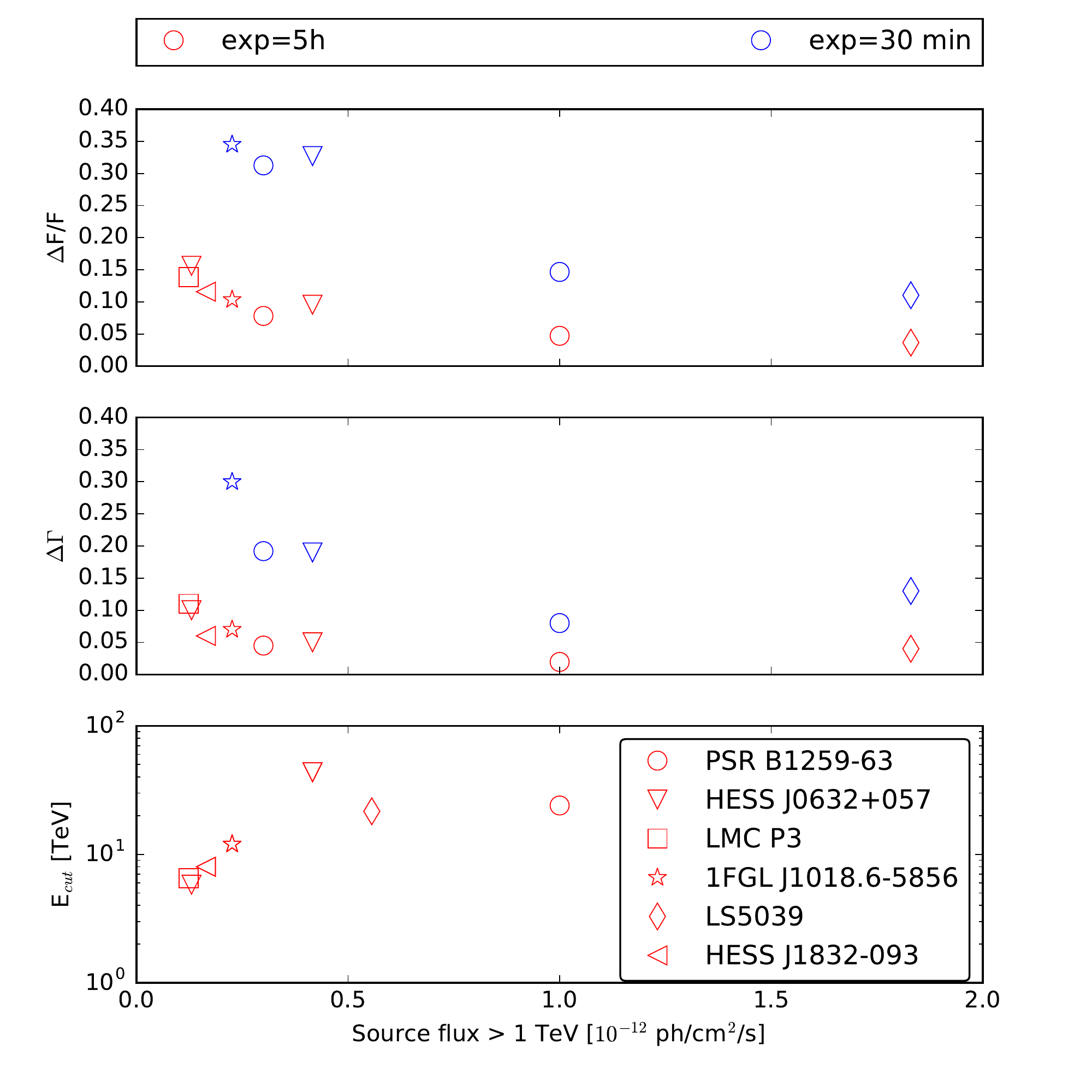}
  \includegraphics[width=0.95\columnwidth]{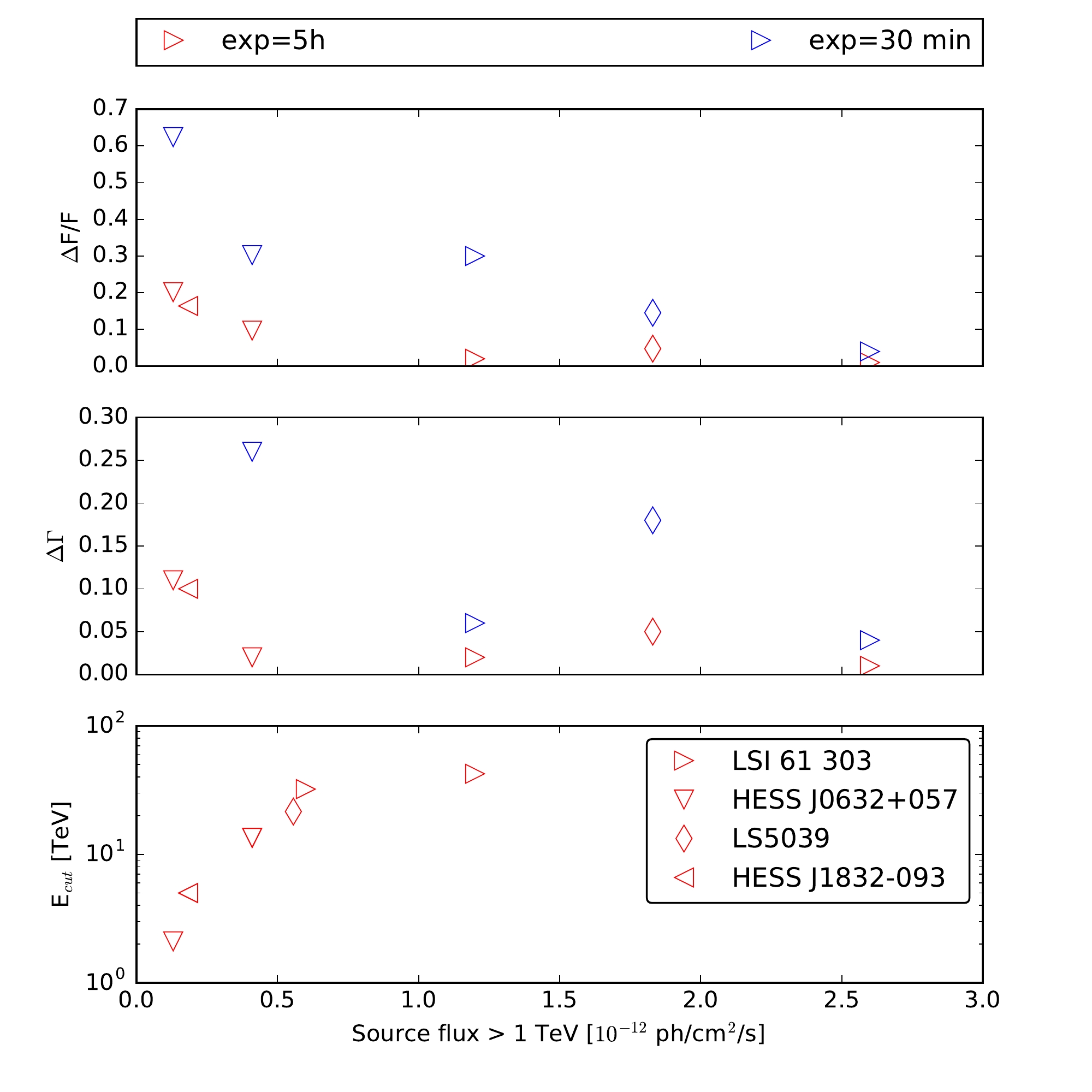}
  \caption{Summary of the simulations performed for the different binaries for various exposure times and telescope configurations. The left and right panels represent the south and north sites, respectively. Exposure time is shown with colours: blue corresponds to 30 min and red to 5 h. In this figure, we show the dependence of the relative flux error (top panel), spectral slope error (middle panel) and maximum energy up to which a cut-off can be excluded (bottom panel).}
  \label{fig:binaries}
\end{figure*}

In order to test the future capabilities of CTA  we have calculated the predicted errors on the spectral parameters for different characteristic fluxes on 30 min and 5 h time scales in the 1 -- 100~TeV energy range. To do this we considered 100 random realisations of the region surrounding the binary. For each simulation we 
used a power-law spectral shape 
and assumed flux and spectral slope values that are typical for the simulated \grb. The uncertainties are defined as a standard deviation of the distribution of best-fit values. The results are shown on the top and middle panels of Fig. \ref{fig:binaries}.
 
While for most systems the spectral shape above 1\,TeV nicely follows a power law, it is not yet clear at which energy we should expect a cut-off. 
In order to estimate the maximum energy up to which CTA will be able to firmly detect a cut-off we fitted event data (simulated for a  power law spectral energy distribution) with a cut-off power law model. This was repeated 1000 times for different data realisation. From the obtained distribution of best-fit cut-off values we found a value above which 95 \% of all cut-off values are located. This corresponds to the 95\% upper limit on the cut-off energy, i.e. if the cut-off is detected  by CTA at 
energies lower than this, 
one can be confident that it is real. 

The resulting 95\% confidence values for sources with fluxes $F(>1\,{\rm TeV}) <1.5 \times 10^{-12}$ ph cm$^{-2}$ s$^{-1}$ are shown in the bottom panel of Fig. \ref{fig:binaries}.
For sources with higher fluxes the resulting value of the cut-off is close to 100\,TeV.  Note that \psrb has a much softer spectrum
($\Gamma\sim$2.9) in comparison with other binaries ($\Gamma\sim$2.3), which results in a lower value of the cut-off energy that can be detected by CTA.

\begin{figure}
  \centering
  \includegraphics[width=\columnwidth]{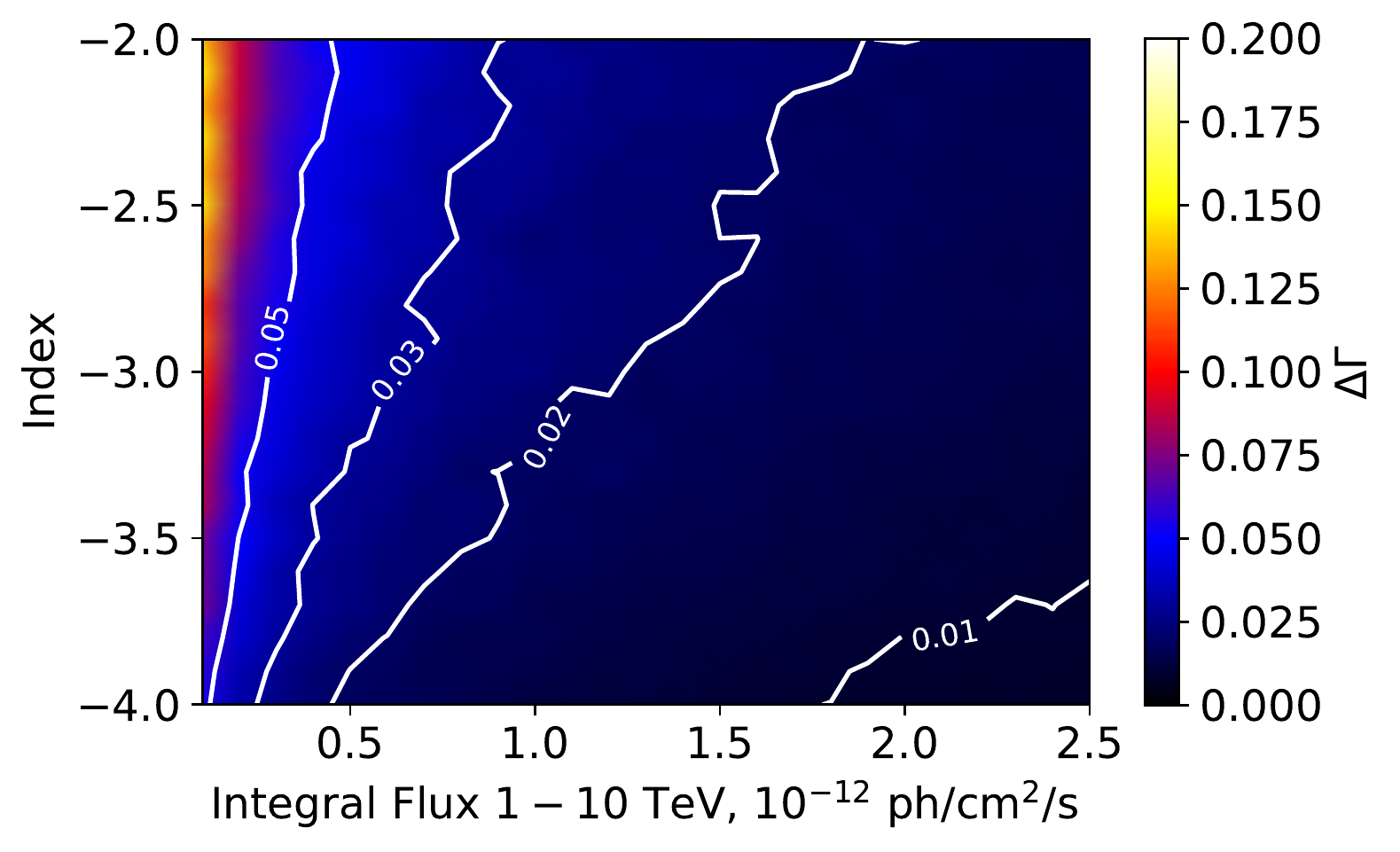}
  \caption{Uncertainty on the slope (colour bar) as a function of the slope value and the flux in the $1-10$~TeV energy range for 5~h observations  of a point-like source at the position of \psrb. White lines illustrate the levels of constant slope uncertainty.}
  \label{fig:idx_flux}
\end{figure}

We have further confirmed that for a given flux and exposure time the error on the slope has a weak dependence on the slope value. Fig.~\ref{fig:idx_flux} illustrates the slope uncertainty (shown with colour) as a function of the slope and the 1--10~TeV flux level for a 5~h observation of the point-like source located at the position of \psrb. 

Lastly in this section we present an overview of what is known about each source listed in Table \ref{parameters}, discuss the questions that can be answered with the new data measured with the precision shown in Fig. \ref{fig:binaries},  and present results of other simulations specific to each individual case.

\subsection{\psrb}
\subsubsection{Source properties}

\object{PSR B1259-63} 
was first discovered as part of a search for short-period pulsars with the Parkes 64-m telescopes \citep{1992MNRAS.255..401J}, and was the first  radio pulsar discovered in orbit around a massive non-degenerate star, the Be star LS 2883 \citep{1992ApJ...387L..37J}. Long-term monitoring of the pulsar has allowed for a very accurate determination of the binary orbit and reveals that \psrb is in a highly eccentric 3.4 years orbit \citep[][and references therein]{2014MNRAS.437.3255S}. 

Radio observations around periastron show an increase and variability in the dispersion measurement of the pulsed signal as the pulsar passes into the stellar wind \citep[e.g.][]{2001MNRAS.326..643J}. This is followed by an eclipse of the pulsed signal from $\sim 16$ days before until $\sim 16$ days after periastron, accompanied by the detection of unpulsed radio emission \citep[][and references therein]{2005MNRAS.358.1069J}. 
The unpulsed emission shows a double peak structure, reaching a maximum around the time of the start and end of the pulsar eclipse, though the shape varies from periastron passage to periastron passage \citep[e.g.][]{2005MNRAS.358.1069J}. The unpulsed emission originates from the extended pulsar wind nebula, which is shown to extend beyond the binary by observations with the Australian Long Baseline Array \citep{2011ApJ...732L..10M}.

The best optical analysis of the optical companion, LS 2883, comes from high resolution spectroscopic observations with the UVES/VLT \citep{2011ApJ...732L..11N}. It is a rapidly rotating O9.5Ve star with an oblate shape and a temperature gradient from the equator to the poles. The star is wider and cooler at the equator ($R_{\rm eq} \approx 9.7\,R_\odot$; $T_{\rm eff,eq} \approx 27\,500$\,K), and narrower and hotter at the poles ($R_{\rm pol} \approx 8.1\,R_\odot$; $T_{\rm eff,pol} \approx 34\,000$\,K).  The Be nature of the star is clear from strong emission lines observed from the source, which originate from the out-flowing circumstellar disk \citep{1992ApJ...387L..37J,1994MNRAS.268..430J,2011ApJ...732L..11N}. The disk is believed to be tilted relative to the orbital plane \citep[e.g.][]{1998MNRAS.298..997W}, with the pulsar crossing the disk plane twice per orbit. Observations have shown that the circumstellar disk is variable around periastron, with the strength of the H$\alpha$ line increasing until after periastron, as well as changes in the symmetry of the double peaked He~{\sc i} line \citep{2014MNRAS.439..432C,chernyakova15, 2016MNRAS.455.3674V}. 

After first being detected at X-ray energies with ROSAT \citep{1994ApJ...427..978C}, observations around periastron have shown a remarkable similarity during different periastron passages. X-ray observations folded over multiple epochs show that the X-ray flux peaks before and after periastron, at around the same time as the pulsed radio emission becomes eclipsed \citep[e.g.][and references therein]{chernyakova15}. This is interpreted as being associated with the time the pulsar passes through the plane of the circumstellar disk. Observations around the 2014 periastron passage also revealed that the rate at which the flux decreased after the second maximum ($\sim 20$\,d after periastron) slowed down and plateaued around 30 days after periastron, at the time when the GeV  $\gamma$-ray emission began to rapidly increase. Extended X-ray emission has also been detected around \psrb, with an extended structure flowing away from the binary; this is suggested to be a part of the circumstellar disk ejected from the system and begin accelerated outwards by the pulsar wind \citep{2011ApJ...730....2P,2015ApJ...806..192P}.

While not detected by COMPTEL and EGRET \citep{1996A&AS..120C.221T}, \psrb has subsequently been detected at GeV and TeV  $\gamma$-ray energies with \flat\ and H.E.S.S.\  The H.E.S.S.\ telescope has reported on observations of the source over the 2004, 2007, 2010 and 2014 periastron passages \citep{2005A&A...442....1A,2009A&A...507..389A,2013A&A...551A..94H,2017arXiv170800895R}. The combined light curves over multiple epochs are beginning to show an indication of a double hump structure around periastron, with a dip at periastron. This is similar to what is observed at X-ray energies. The observations at GeV energies, with \flat, show a very different result. During the 2011 periastron passage, there was a very faint detection around periastron, but then $\sim 30$ days after periastron there was a rapid brightening (flare) with a luminosity approaching that of the pulsar spin-down luminosity \citep{2011ApJ...736L..10T,2011ApJ...736L..11A}. This occurred at a time when the multi-wavelength emission was decreasing and a flare at this period was not expected.   Observations around the following periastron, in 2014, had a substantially shorter exposure before the flare and  no emission was detected before or at periastron. While the flux started to increase at around the same orbital phase, the emission peaked later and was fainter than during 2011 \citep{2015ApJ...798L..26T, 2015ApJ...811...68C}. The most recent periastron passage in 2017 has also shown a different light curve: while no  $\gamma$-ray flare was reported 26-43 days after periastron \citep{2017ATel10918....1Z}, a rapid flare was detected at 70 days after periastron, during a period when GeV emission has previously not been detected \citep{2017ATel11028....1J}. In addition, rapid $\sim3$\,hour flares in GeV and changing UV flux have been reported during the last periastron passage \citep{2018arXiv180409861T}. Detailed analysis of the short time scale variability of the source by \cite{2018ApJ...863...27J} reveal even shorter substructures on a $\sim$ 10 minutes time scale. The energy released during these short flares significantly exceeds  the total spin-down luminosity. This demonstrates a clear  variability of the emission on very different time scales from as short as few minutes, up to orbit-to-orbit variability.


\subsubsection{Prospects for CTA observations}

The TeV  $\gamma$-ray emission from \psrb has been detected from around 100 days before until 100 days after periastron, with the next periastron occurring on 2021 Feb 9. The $\approx3.4$\,yrs orbital period makes observations more challenging as orbit to orbit variation studies must take place over long time periods. Despite this, the improved sensitivity of the CTA observations around the next periastron passages can be used to test different models and better constrain the theoretical models of this source. This can, among others, include: investigating the degree of gamma-gamma absorption around periastron; searching for connections to the GeV flare; and constrain the shape of the light curve near the disk crossings. 

The double hump shape of the TeV light curve around periastron has been attributed to, for example:  more efficient particle acceleration during the disk crossing \citep{2012ApJ...750...70T};  hadronic interactions in the disk \citep{2007Ap&SS.309..253N}; time-dependent adiabatic losses modified by the disk \citep{2011A&A...525A..80K}; and increased gamma-gamma absorption around periastron \citep{2017ApJ...837..175S}. 
Gamma-gamma absorption of the TeV photons should be highest a few days before periastron, and if TeV  $\gamma$-rays are produced near the pulsar location, stellar and disk photons should decrease the flux above 1\,TeV, harden the photon index and vary the low energy cut-off \citep{2017ApJ...837..175S}.  The simulated light curve for \psrb\ is shown in Fig.~\ref{fig:binaries} for 30 min observations.  This measurable limits will help to place better constraints on the level of $\gamma \gamma$ absorption in the system. 

The second question CTA can start to answer is whether or not there is any hint at TeV energies of a connection to the GeV flare. The H.E.S.S. observations around the 2010 periastron passage showed that there was no TeV flare at the time of the {\it Fermi} flare \citep{2013A&A...551A..94H}, and similarly no multi-wavelength flare has been detected. However, X-ray observations around the 2014 periastron passage showed that there was a change in the rate at which the flux decreased \citep{chernyakova15}. Observations with CTA around periastron will allow us to look for a similar effect. This will be an important constraint on the underlying emission mechanism.

Finally, CTA's improved sensitivity will enable a more detailed investigation of the shape of the light curve around the periods of the disk crossings. This is an important comparison to make to models that predict various shapes around these periods, such as \citet{2011A&A...525A..80K,2007Ap&SS.309..253N}.  

\begin{figure}
  \centering
  \includegraphics[width=0.95\columnwidth]{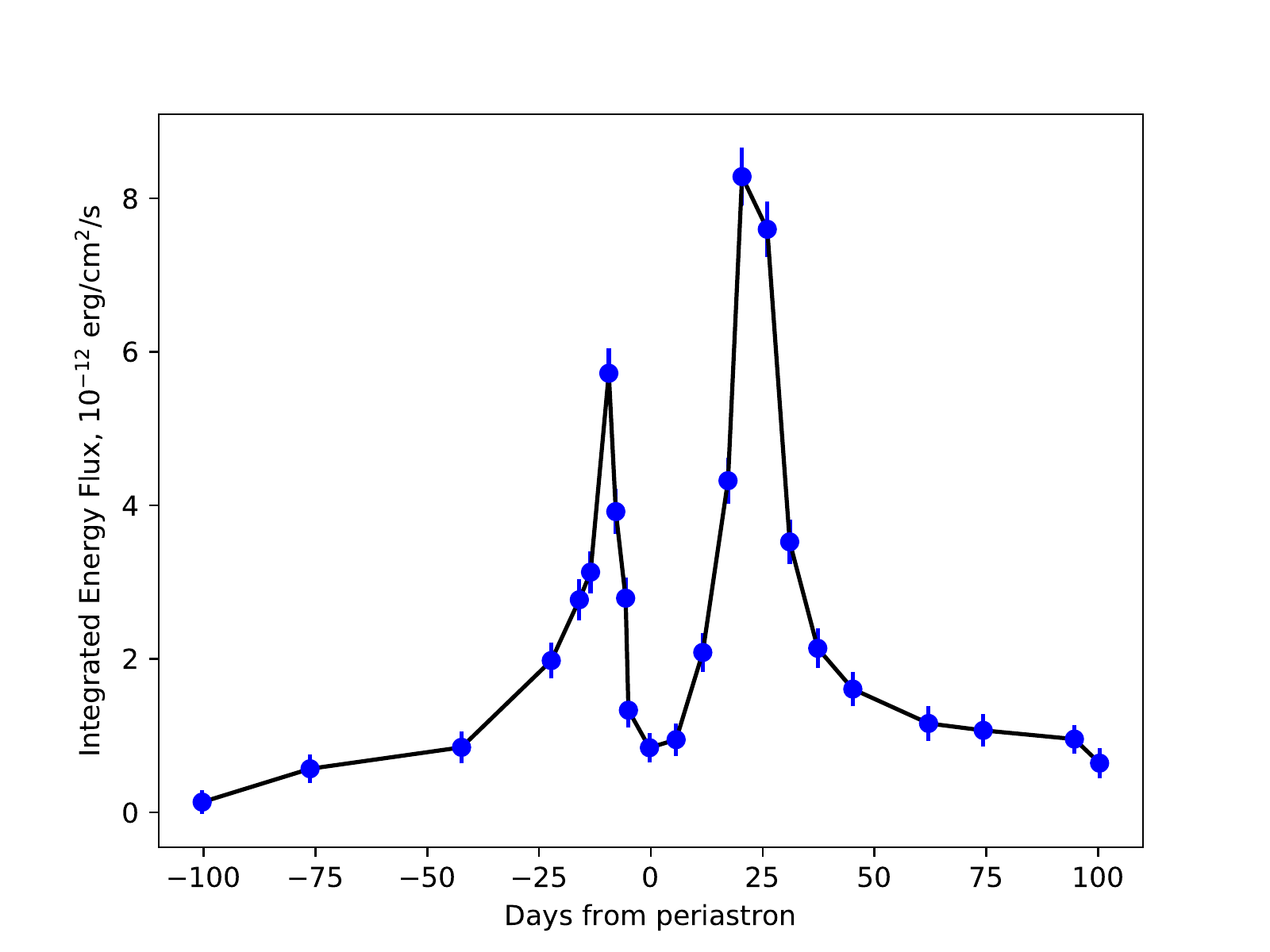}
  \caption{Simulated  light curve above 1~TeV of PSR B1259-63 around the periastron. Each point has a 30 min exposure.}
  \label{psrb:orb}
\end{figure}

To illustrate this point we simulated the light curve of the source around periastron. For this we assumed the constant slope with $\Gamma=2.9$ and modulated the flux above 1 TeV according to the H.E.S.S. observations reported by \cite{2017arXiv170800895R}. Fig. \ref{psrb:orb} illustrates that even 30 min exposures will be enough for CTA to measure the profile with a high accuracy.  

\subsubsection{Source properties}

\object{LS I +61$^\circ$ 303}
was first discovered as a bright  $\gamma$-ray source by the Cos B satellite \citep{1977Natur.269..494H}. Shortly after the discovery, it was realised that this source was also a highly variable radio source  \citep{1978Natur.272..704G} and was associated with the optical source \lsi, a young, rapidly rotating, 10–15 M$_\odot$ B0 Ve star \citep{1979AJ.....84.1030G}. A young pulsar was at first suggested to be responsible for the observed radio emission \citep{Maraschi1981}, but no pulsations have ever been detected, despite intensive searches (e.g. \citealt{2006A&A...459..901S}).

\cite{LSI_2017_BH}  studied the correlation between the X-ray luminosity and the X-ray spectral slope in \lsi and found a good agreement with that of moderate-luminosity black holes. This fact, along with the quasi periodic oscillations observed from this system both in radio and X-rays \citep[e.g.][]{LSI_QPO_2018}, supports a microquasar scenario for \lsi.
A magnetar-like short burst caught from the source, also support the identification of the compact object in \lsi\ with a neutron star \citep{Barthelmy2008, Torres2012}. 
 However, in this case, it will be the only known microquasar which exhibits a regular behaviour, does not demonstrate transitions between various spectral states, and is lacking a spectral break up to hard  $\gamma$-rays.

The radial velocity measurements of the absorption lines
of the primary \citep{2005MNRAS.360.1105C,2009ApJ...698..514A} showed that \lsi is in an elliptical ($e=0.537\pm0.34$) orbit. The orbital period of \lsi\ was found to be $P\approx26.5$~d from radio observations \citep{2002ApJ...575..427G}.
 A strong orbital modulation in \lsi is also observed in the optical/infrared \citep{Mendelson1989,Paredes1994},  X-ray \citep{Paredes1997}, hard X-ray \citep{Zhang2010}, and  HE/VHE $\gamma$-ray \citep{Abdo2009, Albert2009} domains.
In the optical band, the orbital period signature is evident not only in the 
broad band photometry, but also in the spectral properties of the H$_\alpha$ emission line \citep{Zamanov1999}. Due to the uncertainty in the inclination of the system, the nature of the compact object remains unclear, and can be either a neutron star  or a stellar-mass black hole \citep{2005MNRAS.360.1105C}.

In radio, \lsi\ was intensively monitored at GHz frequencies for many years \citep[e.g.][]{1997ApJ...491..381R,2015A&A...575L...9M}. The radio light curve displays periodic outbursts with a position and amplitude changing from one orbit to another. Bayesian analysis of radio data allowed \cite{2002ApJ...575..427G} to establish a super-orbital periodic modulation of the phase and amplitude of these outbursts with a period of $P_{\rm so}=1667\pm 8$ days.
This modulation has also been observed in X rays \citep{2012ApJ...747L..29C,2014ApJ...785L..19L} and $\gamma$~rays \citep{2013ApJ...773L..35A,2016A&A...591A..76A,Xing2017}.

It has been suggested that the super-orbital periodicity can depend  on the Be star disk, either due to a non-axisymmetric structure rotating with a 
period of 1667 days \citep{Xing2017},  or because of a quasi-cyclic build-up and decay of the Be decretion disk \citep{Negueruela-2001,2013ApJ...773L..35A, Chernyakova2017}. Another possible scenario for the super-orbital modulation is related to the precession of the Be star disk \citep{Saha_2016} or  periodic Doppler boosting effects of a precessing jet \citep{2016AA...585A.123M}.

The precessing jet model is based on high resolution radio observations suggesting the presence of a double-sided jet \citep{Massi1993, Paredes1998,Massi2004}. 
The precession period in this model is about 26.9 days, very close to the orbital one.  In this case the observed super-orbital variability is explained as a beat period of the orbital and precession periods \citep{2013A&A...554A.105M}.

\lsi was unambiguously detected  
at GeV energies by \textit{Fermi}-LAT (\citealt{Abdo2009}), thanks to its flux modulation at the orbital period. 
The \flat light curve shows a broader peak after periastron and a smaller peak just before apastron \citep{2014A&A...572A.105J}.
The peak at apastron is affected by the same orbital shift as the radio outbursts and varies on the superorbital time scale, leading to a decline in the orbital flux modulation as the two peaks merge. 

A long-term investigation of \textit{Fermi}-LAT data 
by \citet{Saha_2016} showed  the orbital spectral variability of the  source. The observed spectra is consistent with an exponential cut-off power law with a cut-off at 6--30 GeV for different orbital states of the system. The excess above the spectral cut-off is part of a second emission component dominant at the TeV domain \citep{Hadasch2012,Saha_2016}.

Detected at TeV energies by MAGIC \citep{Albert2006} and by VERITAS \citep{Acciari2008}, the VHE emission from \lsi\ 
shows a modulation consistent with the orbital
period \citep{Albert2009} with the flux peaking at apastron. A decade-long VERITAS observation of \lsi allowed TeV emission to be detect from the system along the entire orbit with the integral flux above 300 GeV varying in the range $(3 - 7) \times 10^{-12} \mathrm{cm}^{-2} \mathrm{s}^{-1}$.
The VHE emission  is well described by a simple power-law spectrum, with a photon index of
$\Gamma=2.63\pm0.06$ near apastron and $\Gamma=2.81\pm0.16$ near periastron \citep{2017ICRC...35..712K}. 

Similar to other wavelengths the TeV 
curve varies from orbit to orbit. 
MAGIC observations during 2009-2010 caught \lsi\ in a low state, with the TeV flux about an order of magnitude lower than was previously detected at the same orbital phase  \citep{2012ApJ...746...80A}.

Long term multi wavelength monitoring of \lsi indicates a correlation between the X-ray ({\it XMM-Newton}, {\it Swift}/XRT) and TeV (MAGIC, VERITAS) data sets. 
At the same time GeV emission shows no correlation with the TeV emission which, along with the spectral cut-off at GeV energies, implies that the GeV
and TeV gamma rays originate from different particle populations   \citep{Anderhub2009,2013ApJ...779...88A,2015ICRC...34..818K}.

\subsubsection{Prospects for CTA observations}

Correlation of VHE and X-ray emission might indicate that in this source the synchrotron emission visible at X-rays is due to the same electrons that produce the TeV emission by inverse Compton scattering of stellar photons. 
However, while X-ray variability on time scales of thousands of seconds is known from the source \citep{2006A&A...459..901S}, MAGIC and VERITAS observations require much longer exposure times, making it difficult to clearly compare the spectral behaviour at different energies. CTA's sensitivity is crucial for detecting spectral variability on comparable time scales at both X-rays and VHE energies.

We studied the capabilities of CTA to unambiguously detect  spectral variability  of \lsi on different time scales by performing a series of simulations, based on  existing observations.
It has been observed that the spectrum of the TeV emission varies  between the low and the high state. For our simulations we chose ${\rm F(E>1~TeV)} = 2.6\times10^{-12}$~cm$^{-2}$~s$^{-1}$ for the high state and ${\rm F(E>1~TeV)} = 1.2\times10^{-12}$~cm$^{-2}$~s$^{-1}$ for the low state. We assumed a power-law model to describe the source and used different spectral slopes, in order to see whether CTA would be able to distinguish between them.  
The spectral slopes that were chosen were $\Gamma =$ 2.4, 2.7 and 3.0, in agreement with MAGIC and VERITAS observations.
Simulations of both 30 min and 5 h exposure times where performed for each set of parameters. 
Each combination was simulated 500 times to ensure enough statistics.
In the analysis of each realisation, the normalisation and spectral index were kept as free parameters. 

\begin{figure}
  \centering
  \includegraphics[width=0.95\columnwidth]{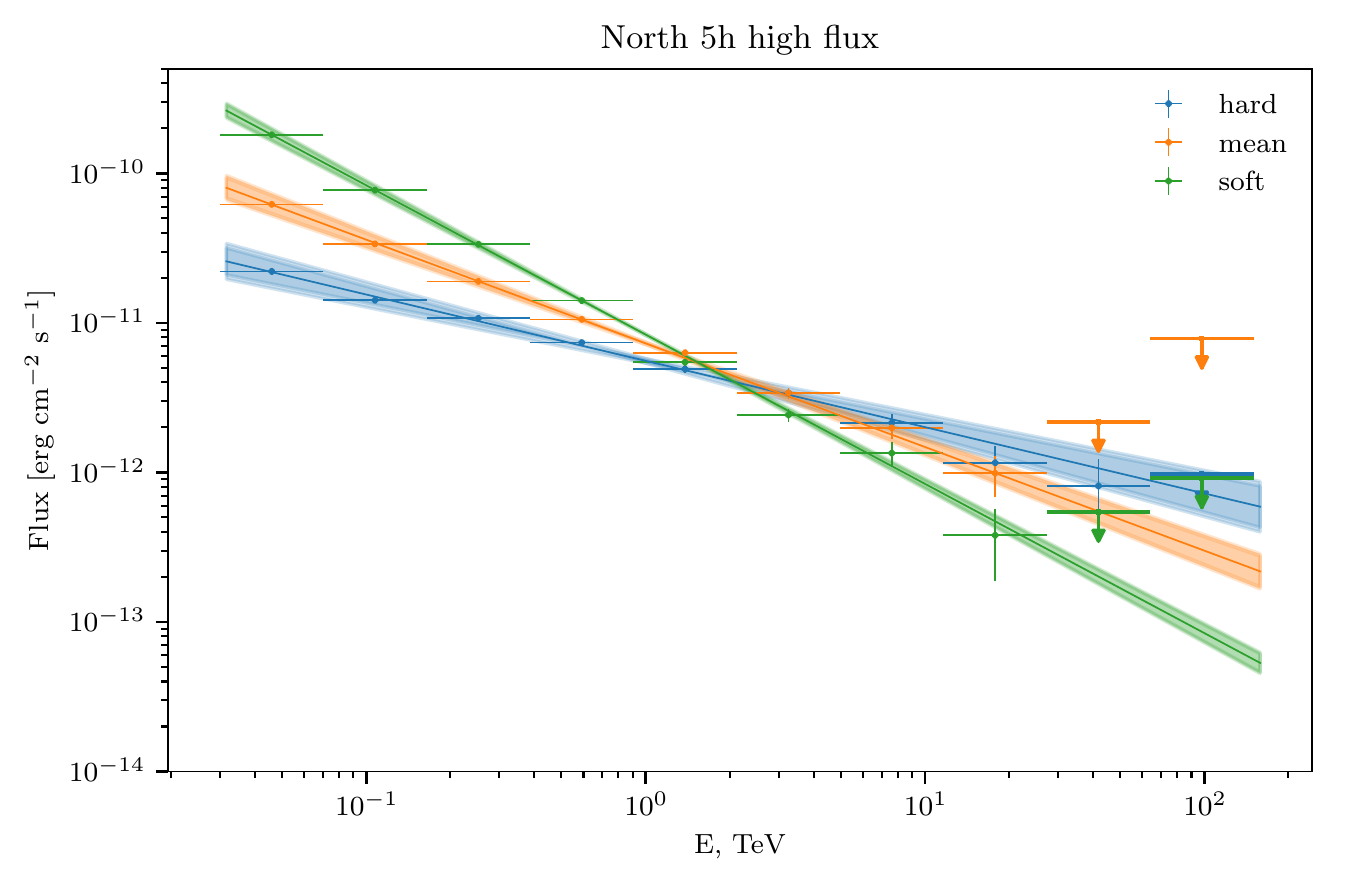}
  \caption{\lsi simulated spectra for 5~h observation. The three different spectral slopes are shown. Solid lines represent the power-law spectral fit. Butterfly represents $1\sigma$ uncertainty in the fit.}
  \label{fig:lsi_spectra}
\end{figure}

Similarly to the simulations presented in Fig. \ref{fig:idx_flux}, we found that the uncertainty on the spectral slope has a weak dependence on the slope value and a 5~h observation is enough to determine the slope with an accuracy of better than 0.1 (see Figs \ref{fig:binaries}, \ref{fig:lsi_spectra}).

To check the ability of CTA to detect a VHE cut-off in \lsi, we simulated a 5~h observation with a power-law spectral model and fit the data with an exponential cut-off power law. The resulting values are shown in Fig. \ref{fig:binaries}.

\begin{figure}
  \centering
  \includegraphics[width=0.95\columnwidth]{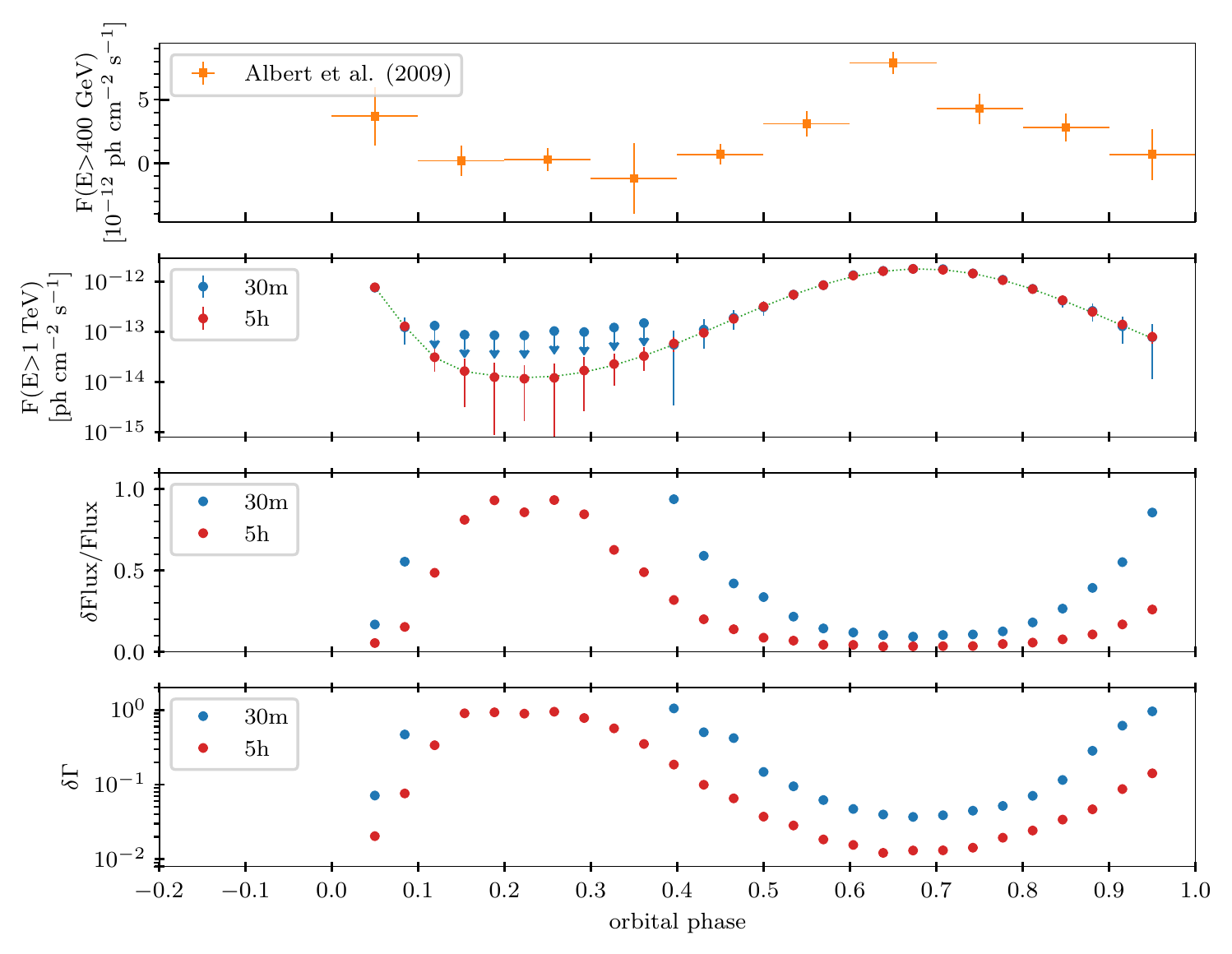}
  \caption{
  The upper panel shows the original orbital light curve of \lsi as observed by MAGIC \citep{Albert2009}; the next panel is the simulated orbital light curve at ${\rm E > 1~TeV}$, where the green line shows the simulated flux and the arrows represent 2 $\sigma$ upper limits; the third panel shows the relative uncertainty of the simulated flux; and the  bottom panel shows the uncertainty in the simulated photon index. In all panels exposure time is shown with colours: blue corresponds to 30 min and red to 5 h.}
  \label{fig:lsi_orbital}
\end{figure}

Finally, we performed a study of the orbital variability of the source. 
We took the light curve above 400~GeV obtained by \cite{Albert2009} and modelled it with 27 bins (see Fig. \ref{fig:lsi_orbital}), since we wanted to study the inter-night variability of the source. 
We simulated the source with a power-law spectrum with a photon index $\Gamma = 2.7$. In the reconstruction photon index, as well as the normalisation,  were left free to vary.
We performed 100 realisations for each orbital bin for 30~min and 5~h exposures. In this analysis we assumed much lower value for the flux in the low state (orbital phases between 0.2 and 0.4) of about $10^{-14}$~cm$^{-2}$~s$^{-1}$. The resulting uncertainties for the relative flux and slope  are summarised in Fig. \ref{fig:lsi_orbital}. All uncertainties are statistical only at a $1~\sigma$ confidence level and are below 10$\%$ for the integral flux in the high state  with the photon index uncertainty below $0.1$ even for a 30 min exposure. In the low state the source is barely detected even with a 5 h exposure. The upper limits shown correspond to the 2$\sigma$ confidence level. This simulation shows that if the flux of the source is above $\sim 10^{-13}$~cm$^{-2}$~s$^{-1}$ CTA will be able to detect inter-night variability of the source at a 10\% level. Such precision will  allow the superorbital variability of the orbital profile to be studied and compared to other energy bands. In the high state it will be possible to study the variability of the source at a 30 min time scale, comparable to what is observed in X-ray data \citep[see e.g.][]{Chernyakova2017}.

\subsection{\ls \label{section:ls}}

\subsubsection{Source properties}

\object{LS 5039}
has the shortest orbital period thus far of all known  $\gamma$-ray binaries (3.9\,d, see Table \ref{parameters}). Also known as V497 Sct, based on {\it ROSAT} X-ray data, \citet{1997A&A...323..853M} first reported it as a high-mass X-ray binary. Its peculiar nature as a persistent non-thermal radio emitter was  soon revealed after the detection of a bright radio counterpart with the Very Large Array (VLA) by \citet{1998A&A...338L..71M}. This fact already anticipated the capability of the system to accelerate 
electrons to relativistic speeds. Follow-up Very Long Baseline Interferometry (VLBI) images resolved the radio emission into elongated features, 
and as a result \ls\ was interpreted as new microquasar system \citep{2000Sci...288.2340P}. Moreover, at the same time it was also tentatively associated with the EGRET $\gamma$-ray source 3EG J1824$-$1514. The confirmation of \ls\ as an unambiguous (>100 GeV) $\gamma$-ray source was finally obtained with H.E.S.S. \citep{2005Sci...309..746A}.

During the twenty years since its discovery, the physical picture of \ls\ has generally evolved from the microquasar scenario to a binary system hosting a young non-accreting neutron star  interacting with the wind of a massive O-type stellar companion (see e.g. \citealt{2013A&ARv..21...64D} and references therein). This is strongly supported by VLBI observations of periodic changes in the radio morphology \citep{2012A&A...548A.103M}, although no radio pulsations have been reported so far.

At different photon energies, the shape of the \ls light curve varies, as confirmed in the most recent multi-wavelength studies using {\it Suzaku}, {\it INTEGRAL}, {\it COMPTEL}, {\it Fermi}-LAT and H.E.S.S. data \citep[][and references therein]{2016MNRAS.463..495C}. The X-ray, soft $\gamma$-ray (up to 70 MeV) and TeV emission peak around 
inferior conjunction after the apastron passage. In contrast, $\gamma$-rays in the 0.1-3 GeV energy range anti-correlate and have a peak near the superior conjunction soon after the periastron passage. No clear orbital modulation is apparent in the 3-20 GeV band. This dichotomy suggests the existence of a highly relativistic particle population accounting for both X-ray/soft $\gamma$-ray and TeV emission mainly by synchrotron and anisotropic inverse Compton (IC) scattering of stellar photons, respectively. The GeV $\gamma$-ray peak would arise when TeV photons (of an IC origin) are absorbed  through pair production as the neutron star gets close to its O-type companion, and further enhances the GeV emission through cascading effects. Variable adiabatic cooling and Doppler boosting are other effects proposed to play an important role when trying to understand the multi-wavelength modulation of systems like \ls\ \citep[see for instance][]{2008IJMPD..17.1909K, Suzaku2009, 2013A&ARv..21...64D}.

\subsubsection{Prospects for CTA observations}

In order to estimate CTA capabilities to detect the temporal and spectral variations of 
emission from \ls, we have simulated CTA observations of this source at different spectral states.

Since the emission spectrum of \ls\ 
varies with orbital phase, we have assumed the flux and spectral shape modulations found in the recent H.E.S.S. data~\citep{Mariaud15}. These observations suggest the source spectrum varies from $dN/dE \sim E^{-1.9} \exp{(-E/6.6\mathrm{TeV})}$ 
at inferior conjunction (phase $\approx 0.7$) to $dN/dE \sim E^{-2.4}$ at superior conjunction (phase $\approx 0.05$).
We have further assumed that the source spectrum always follows a power law with an exponential cut-off shape and have taken the flux and spectral index evolution from Figs.~3 and 4 of \citet{Mariaud15}; as no spectral cut-off was observed at superior conjunction, we have assumed that the cut-off energy is modulated between $E_{\rm cut}^{\rm min} = 6.6$~TeV at phase 
$\phi \approx 0.7$ and $E_{\rm cut}^{\rm max} = 40$~TeV at phase $\phi \approx 0.3$:
  \begin{equation}
    \log_{10}(E_{\rm cut}) = \log_{10}(E_{\rm mean}) - \Delta \log_{10}E \times \cos(\phi-0.71)
  \end{equation}
where $\log_{10}(E_{\rm mean}) = 0.5\times[\log_{10}(E_{\rm cut}^{\rm max}) + \log_{10}(E_{\rm cut}^{\rm min})]$ and $\Delta \log_{10}E =  0.5\times[\log_{10}(E_{\rm cut}^{\rm max}) - \log_{10}(E_{\rm cut}^{\rm min})]$.

\begin{figure}
  \centering
  \includegraphics[width=1.0\columnwidth]{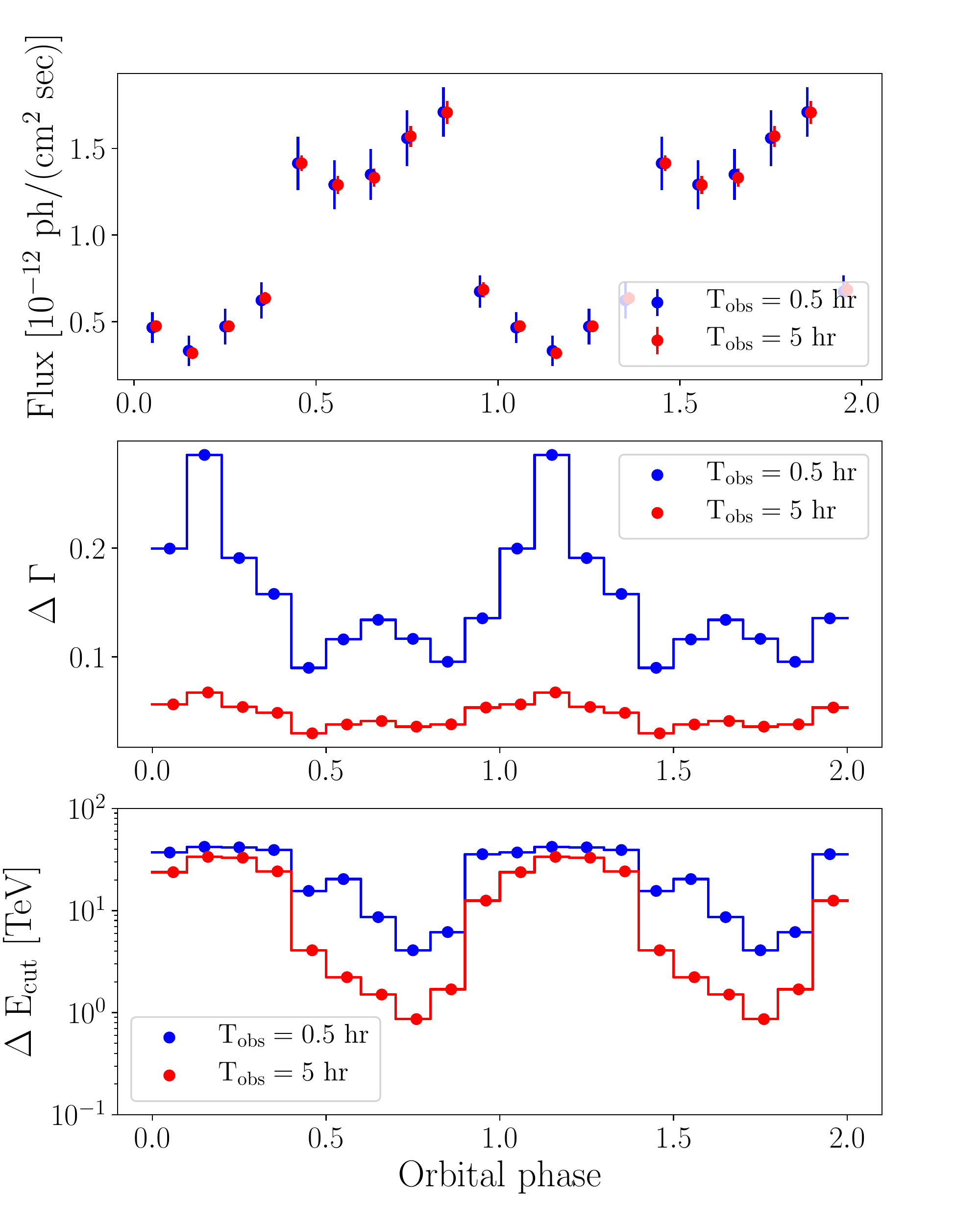}
  \caption{Simulated CTA view of \ls orbital variations above 1~TeV energy for 0.5 and 5~hr long snapshots as observed from the southern CTA site. The upper panel gives the integral flux above 1~TeV, and the  middle and lower panels the estimated uncertainty of the measured spectral index and cut off energy, correspondingly. The data points corresponding to $T_{obs}=5$~hr are shifted to the right by 0.01 phase to improve the readability of the plot.}
  \label{fig:ls5039_orbital_light_curve}
\end{figure}

We have simulated 10 snapshot observations 
from orbital phases $0.0$ to $1.0$, lasting 0.5~hr and 5~hr each. This 
gives a total exposure of 5 and 50~hr on the source. To reconstruct the simulated flux, we have assumed the same power-law with exponential cut-off model, however with 
the spectrum normalisation, index and cutoff energy as free parameters. For each phase bin and observation duration, the simulation was repeated 100 times to estimate the mean values of the flux (in the 1-100~TeV range), spectral index and cut-off energy, as well as their standard deviations. The results of these simulations are shown in Fig.~\ref{fig:ls5039_orbital_light_curve}; the estimated uncertainties of the reconstructed source spectral parameters are also given there.

As can be seen from Fig.~\ref{fig:ls5039_orbital_light_curve}, CTA can follow the orbital flux evolution of \ls even with 30~minute observational snapshots. However in the orbital phase range $0.1-0.3$ the uncertainties on the flux become $\gtrsim 20\%$ and at least 5~hr long exposures would be required to determine the flux accurately. Such exposure times yield $\lesssim 10\%$ accuracy of the flux and spectral index determination in this phase range, whereas the cut-off energy cannot be measured accurately. Only during the bright flux period, corresponding to the phase range $\sim 0.4-0.9$, is the uncertainty in the cut-off energy better than $\sim 20\%$. This implies, that detailed spectral studies of this particular binary phase 
will require integration over several orbital periods.

Such CTA observations of \ls during its high-flux periods will enable spectral studies on time scales as short as 0.01 orbital periods ($\approx 1$~hr), that will strongly constrain the physical processes at work.

Further, it will clarify the feasibility of the rotating hollow cone model~\citep{2008ApJ...672L.123N}, previously proposed to explain the \ls\ TeV light curve. In this model, the TeV peak is composed of two narrower peaks, whose appearance depends on the location of the emission region and the system geometry. \citet{2008ApJ...672L.123N} suggested that the flux difference between the peaks and inferior conjunction at phase $\sim 0.7$ is $\gtrsim 10\%$. Such variation are detectable with CTA in several hours of exposure, as one can see from the top panel of Fig.~\ref{fig:ls5039_orbital_light_curve}. In this way CTA observations may allow the geometry of the system to be constrained, including the otherwise elusive orbital inclination angle.

\subsection{\fgl}

\subsubsection{Source properties}

\object{1FGL J1018.6-5856}
(3FGL J1018.9-5856, HESS J1018-589A) is a point-like, high-energy $\gamma$-ray source, positionally coincident with the supernova remnant SNR G284.3–1.8. Using the Gaia DR2 source parallax and assuming a Gaussian probability distribution for the parallax measurement, \citet{Marcote+18} derived a source distance of $d = 6.4^{+1.7}_{-0.7}$ kpc. They also calculated the Galactic proper motion of the source and found that it is moving away from the Galactic plane. Both the source distance and proper motion are not compatible with the position of the SNR G284.3-1.8 (which is located at an estimated distance of $\simeq$ 2.9 kpc). Therefore, it is possible to exclude any physical relation between the binary source and the SNR.

Spectroscopic observations of the optical counterpart allowed  \citet{Strader+15}  to find that a companion star has a low radial velocity semi-amplitude of 11-12 km s$^{-1}$, which favours a neutron star as a compact object. This conclusion is in agreement with the results of \citet{Monageng+17}, who constrained the eccentricity of the orbit $e = 0.31 \pm 0.16$ and showed that the compact object is a neutron star, unless the system has a low inclination $i \lsim 26^\circ$.

The 1FGL J1018.6–5856 was detected in a blind search for periodic sources in the \flat survey of the Galactic Plane. Optical observations show that the non-thermal
source is positionally coincident with a massive star of spectral
type O6V(f). 
The radio and X-ray fluxes from the source are
modulated with the same period of 16.544 days, interpreted
as the binary orbital period \citep{LAT2012}.

The very-high-energy counterpart of this source is the point-like source HESS J1018-589A, \citep{HESS2012}. Thanks to a dedicated observation campaign at VHE, HESS J1018-589A was detected up to 20 TeV. Its energy spectrum is well described with a power-law model, with a photon index $\Gamma$ = 2.2 and a mean differential flux  $N_{0}$ = (2.9$\pm$0.4)$\times 10^{-13}$ ph cm$^{-2}$ s$^{-1}$ TeV$^{-1}$ at 1 TeV. As in the case of other $\gamma$-ray binaries, the VHE spectrum cannot be extrapolated from the HE one, which has a break  at around 1~GeV. The orbital light curve at VHE peaks in phase with the  X-ray and HE (1-10 GeV) ones.

Based on optical spectroscopic observations, \citet{Strader+15} found that the maxima of the X-ray, HE and VHE flux correspond to the inferior conjunction.  This finding was unexpected, since $\gamma$-rays are believed to be produced via anisotropic inverse Compton up-scattering of the stellar UV photons. Therefore, the peak of the $\gamma$-ray flux should occur at the superior conjunction,  especially if the system is edge-on. This discrepancy could be explained only if the binary orbit is eccentric and the flux maximum occurs at the periastron.

 {NuSTAR observations \citep{An+15} demonstrated that, similar to other $\gamma$-ray binaries, the broad band X-ray spectrum is well fitted with an unbroken power-law model.} 
The source flux shows a correlation with the spectral hardness throughout all orbital phases.

A comparison of the light curves of \fgl\ at different energy ranges shows that both the X-ray and the low-energy (E $<$ 0.4 GeV) $\gamma$-ray bands are characterised by a similar modulation (a broad maximum at $\phi$ = 0.2--0.7 and a sharp spike at $\phi$ = 0), thus suggesting that they are due to a common spectral component. On the other hand, above $\sim$ 1 GeV the orbital light curve changes significantly, since the broad hump disappears and the remaining structure is similar to the light curve observed at VHE. Based on these results, \citet{AnRomani17} suggested that the flux in the GeV band is due mainly to the pulsar magnetosphere, while the X-ray flux is due to synchrotron emission from shock-accelerated electrons and the TeV light curve is dominated by the up-scattering of the stellar and synchrotron photons, via External Compton (EC) and Synchrotron-Self Compton (SSC) mechanisms, in an intrabinary shock.  
The light curves at different energy ranges can be reproduced with the beamed SSC radiation from adiabatically accelerated plasma in the shocked pulsar wind. 
This is composed of a slow one and a fast outflow. Both components contribute to the synchrotron emission observed from the X-ray to the low-energy $\gamma$-ray band, which has a sinusoidal modulation with a broad peak around the orbit periastron at $\phi$ = 0.4. On the other hand, only the Doppler-boosted component reaches energies above 1 GeV, which are characterised by the sharp maximum which occurs at the inferior conjunction at $\phi$ = 0.
This result can be obtained with an orbital inclination of $\sim$ 50$^{\circ}$ and an orbital eccentricity of $\sim$ 0.35, consistent with the constraints obtained from optical observations. In this way the model could also explain the variable X-ray spike coincident with the $\gamma$-ray maximum at $\phi$ = 0.

\subsubsection{Prospects for CTA observations}
Although \fgl\ was investigated in depth over the last few years, several issues about its properties are still pending, such as 
the physical processes which produce the HE/VHE emission. Moreover, it is still not clear whether the X-ray and $\gamma$-ray peaks are physically related to the conjunctions or the apastron/periastron passages. Therefore, the observation of this source with CTA will allow us to address a few topics. Thanks to the high sensitivity of CTA it will be possible to investigate the orbital modulation of the source spectrum and to study the correlation of the VHE emission with the system geometry. From the spectral point of view, the spectral shape will be further constrained at both the low (E $<$ 0.1 TeV) and the high (E $>$ 20 TeV) energy end. This will provide further constraints on the location, magnetic field, and acceleration efficiency of the VHE emitter \citep{LS5039_2008_Khan} and on the opacity due to pair production \citep{BoettcherDermer05,Dubus06}.

To study CTA capabilities we first simulated the phase-averaged spectrum of the source based on the H.E.S.S.\ observations \citep{2015A&A...577A.131H}. As input we assumed a simple power-law emission with a photon index $\Gamma=2.2$ and a flux normalisation $N_0 = 2.9 \times 10^{-19}$ $\mathrm{MeV^{-1}~cm^{-2}~s^{-1}}$ at 1 TeV. We performed three sets of simulations for  30 min, 5h and 50h observations. For each set we performed 100 simulations.
Fig.~\ref{fig:binaries} shows that with a 5 h observation it will be possible to measure the source flux and spectral slope with an uncertainty of $\simeq$ 10 \% and 0.05, respectively. With only 30 min of observation the corresponding errors would be $\simeq$ 35 \% and 0.3.

\begin{figure}[h]
\includegraphics[width=\columnwidth,angle=0]{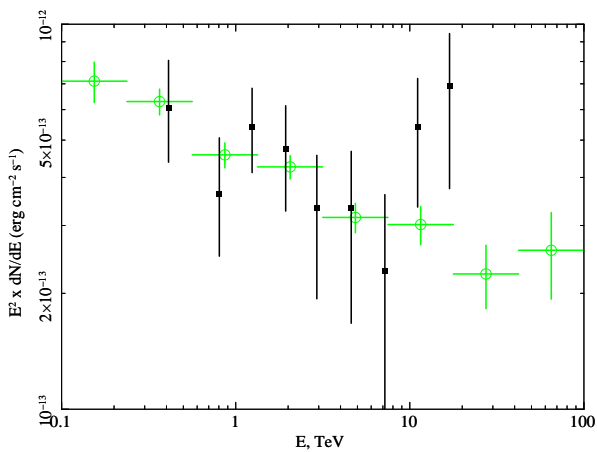}
\caption{\textit{Black}: spectrum of \fgl\ obtained with 63 hours of observation with H.E.S.S.\ \citep{2015A&A...577A.131H}. \textit{Green}: simulation of the source spectrum obtained with 50 hours of observation with CTA. \label{fig:1FGL_spectrum}}
\end{figure}

In Fig.~\ref{fig:1FGL_spectrum} we report the simulated spectrum of \fgl\ obtained with 50 h of observation, together with that obtained with 63 hours of H.E.S.S.\ observations. It shows that the CTA spectrum is well determined both at low energies (down to E $\simeq$ 0.1 TeV) and at high energies (up to E $\simeq$ 100 TeV), thus providing a significant enlargement of the spectral coverage compared to H.E.S.S.. The extension of the spectral range towards the low energies will enable the investigation of the connection to the MeV-GeV emission, while the increase of the high-energy end will be important to constrain the cut-off linked to particle acceleration. In fact, we estimated that, with a 5 h observation, CTA will allow us to detect a high-energy cut-off if it is located below 18 TeV (Fig.~\ref{fig:binaries}).

\begin{figure}[h]
\includegraphics[width=\columnwidth,angle=0]{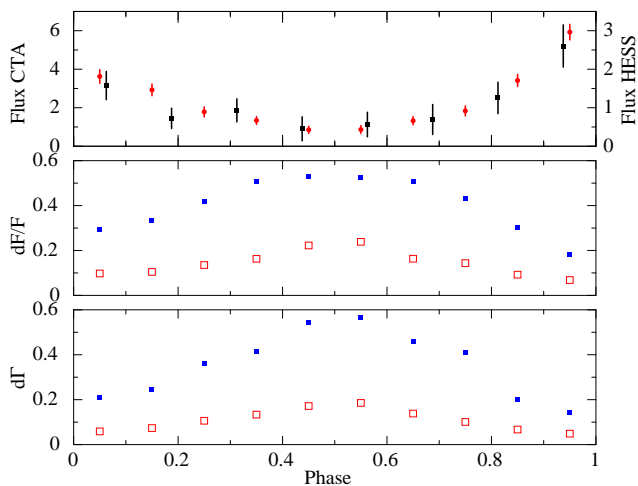}
\caption{\emph{Upper panel}: flux modulation of \fgl\ along the orbital phase as observed with H.E.S.S.\ (black symbols, in units of 10$^{-12}$ ph cm$^{-2}$ s$^{-1}$ for E $>$ 0.35 TeV) and simulated for 5 h of observation with CTA (red symbols, in units of 10$^{-13}$ ph cm$^{-2}$ s$^{-1}$ for E $>$ 1 TeV). \emph{Middle panel}: flux relative error in the case of simulated CTA observations of 30 min (blue filled squares) and 5 h (red open squares). \emph{Lower panel}: uncertainty of the photon index $\Gamma$ in the same cases. \label{fig:1FGL_variability}}
\end{figure}

We also carried out a study on the flux modulation of the source along the orbit. Following  \citet{2015A&A...577A.131H} we divided the whole 
orbit into 10 phase bins
(1 bin $\simeq$ 39.67 h) and  assumed a simple photon spectrum with a power-law model with a slope $\Gamma=2.2$. 
For each phase bin we performed 100 simulations for  both 30min and 5h  observations. We fitted the spectrum with a simple power-law model, keeping both the normalisation and the photon index $\Gamma$ free to vary. The results of this set of simulations are reported in Fig.~\ref{fig:1FGL_variability}, where we show the variability (along the orbital phase) of the 
flux, its relative error, and the uncertainty of the index.

In the upper panel of Fig.~\ref{fig:1FGL_variability}, we report as red symbols the simulated light curve obtained with a campaign of 5 h observations for each phase bin with CTA. It proves that in this case CTA can clearly resolve the source flux variability along the orbit. It will be possible to point out flux variations of $\simeq$ 25 \% over time scales of $\sim$ 0.1 orbital periods (middle panel) and, even at its flux minimum, the source will be detected with a significance $\gsim$ 7 $\sigma$. For comparison, in the upper panel of Fig.~\ref{fig:1FGL_variability} we report as black symbols the folded light curve obtained with H.E.S.S.\ (with $\simeq$ 8 h of observation for each phase bin). We note that, in this case, it is possible to claim a clear flux variability only in the three phase bins in the phase range $\phi$ = 0.8-1.1, while the flux values measured in the remaining phase range are consistent with each other. Thanks to the better characterisation of the source variability provided by CTA, it will be possible to improve the correlation with the X-ray and HE variability and to put tighter constraints on the position and size of the VHE emitter.

The lower panel of Fig.~\ref{fig:1FGL_variability} shows that, even at the flux minimum, 5 h of observation with CTA will provide a measurement of the photon index with an uncertainty of 0.2 ($\simeq$ 9 \%). This accuracy is comparable to that obtained with more than 60 h of observation with H.E.S.S.. Therefore, it will be feasible to point out possible spectral variations $>$ 10 \% along the orbital phase. In this way, it will be possible to single out the VHE emission and absorption processes and to obtain useful information on both the source magnetic field and the efficiency of the particle acceleration and pair production.

\subsection{\hess0632}

\subsubsection{Source properties}
Contrary to other $\gamma$-ray loud binaries, 
\object{HESS J0632+057} 
for a while remained the only system missed in the GeV energy band. Only recently hints of GeV-detection with \fermi-LAT were reported in \citet{malyshev16} and \citet{li17}. The system was initially discovered during H.E.S.S.\ observations of the Monoceros region~\citep{hess_j0632} as an unidentified point-like source. Its spatial coincidence with the Be star MWC~148 suggested its binary nature~\citep{hess_j0632,hinton09}. 
With dedicated observational campaigns, the binary nature of the system was confirmed by radio~\citep{skilton09} and soft X-ray~\citep{falcone10} observations. In the TeV band, the system was also detected by VERITAS and MAGIC~\citep{magic12,Aliu2014HESS0632}.

The orbital period of \hess0632 of $\sim 316 \pm 2$~d~\citep{malyshev17}, with a zero-phase time $T_0 = 54857$~MJD~\citep{bongiorno11}, was derived from \swift/XRT observations. The exact orbital solution and even the orbital phase of periastron is not firmly established and is placed at orbital phases $\phi\sim 0.97$~\citep{casares12} or $\phi \sim 0.4-0.5$~\citep{moritani18,malyshev17}.

The orbital-folded X-ray light-curve of \hess0632 demonstrates two clear peaks of the emission -- first at phase $\phi\sim 0.2-0.4$ and second at $\phi\sim 0.6-0.8$ separated by a deep minimum at $\phi\sim 0.4-0.5$~\citep{bongiorno11,Aliu2014HESS0632}. A low-intermediate state is present at $\phi\sim 0.8-0.2$. The orbital light curve in the TeV energy range shows a similar structure, as was reported by~\citet{veritas15}. Hints of orbital variability in the GeV range were reported in~\citet{li17}.

\begin{figure}
\includegraphics[width=1.05\linewidth]{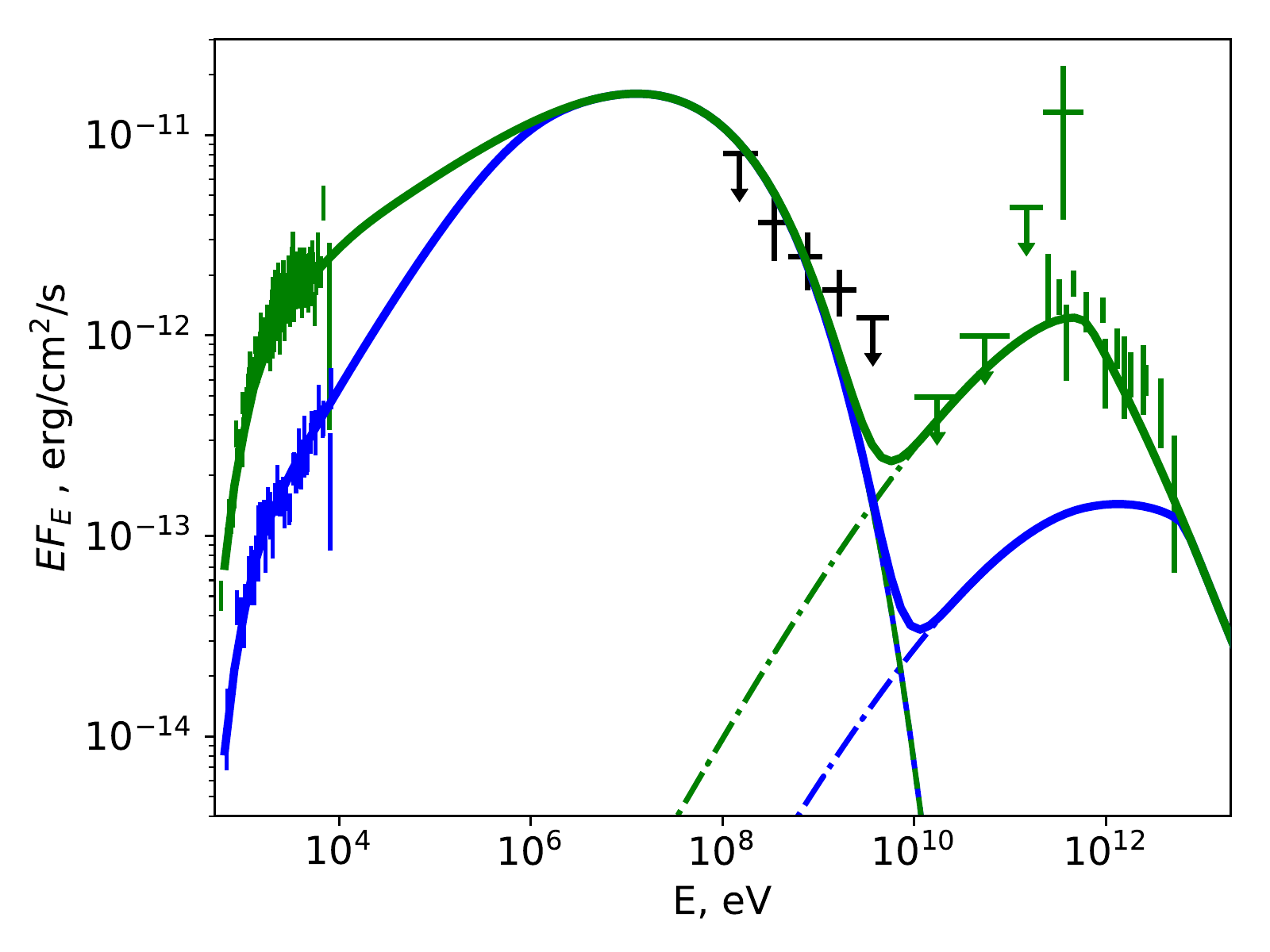}
\caption{The X-ray to TeV spectrum of \hess0632 during its high state (green points; orbital phases $\phi\sim 0.3-0.4$) and low state (blue points; $\phi\sim 0.4-0.5$). The data are adopted from~\citet{malyshev17} (X-rays), ~\citet{li17} (mean GeV spectrum, black points), and~\citet{malyshev16} (green upper limits). TeV data are adopted from~\citet{veritas15}. The solid lines show ``similar to \psrb'' model flux, while dashed and dot-dashed lines illustrate contributions from synchrotron and IC model components correspondingly, see text for more details.}
\label{fig:model}
\end{figure}

The X-ray to TeV spectrum of \hess0632 is shown in Fig.~\ref{fig:model}. Several models were so far  proposed to explain the observed variations of the flux and spectrum along the orbit. In the flip-flop scenario \citep[see e.g.][and references therein]{moritani15} the compact object is assumed to be a pulsar passing a periastron at $\phi=0.97$. Close to the apastron (orbital phases $\sim 0.4-0.6$), the pulsar is in a rotationally powered regime, while it switches into a propeller regime when approaching the periastron (phases $0.1-0.4$ and $0.6-0.85$). In a flip-flop system, if the gas pressure of the Be disk overcomes the pulsar-wind ram pressure, the pulsar wind is quenched (phases $0-0.1$ and $0.85-1$). Because the Be disk of the system is estimated to be about three times larger than the binary separation at periastron, the compact object enters a dense region of the disk near the periastron. In such a situation, the strong gas pressure is likely to quench the pulsar wind and suppress high-energy emissions.
Alternatively, the observed orbital variations can be explained within the ``similar to \psrb'' model~\citep{malyshev17}. The similar two-peak behaviour of the \hess0632 and \psrb orbital light-curves allows us to assume that the orbital plane of \hess0632 is inclined with respect to the disk plane, similarly to \psrb. Orbital X-ray/TeV peaks within this model correspond to the first and second crossing of the disk by a compact object. Higher ambient density during these episodes leads to more effective cooling of the relativistic electrons by synchrotron/inverse Compton mechanisms, resulting in an increased level of X-ray/TeV emission. Note, that the orbital phase of periastron in this model is located at a phase $\phi\sim 0.4 - 0.5$~\citep{malyshev17}.

The break in the GeV-TeV spectrum at $\sim 200$~GeV can be interpreted as a corresponding break in the spectrum of emitting relativistic electrons. The X-ray-to-GeV and TeV parts of the spectrum are explained as synchrotron and IC components. An initial power-law ($\Gamma_{1,e}\sim 1.3$) spectrum of electrons can be modified by synchrotron energy losses at above $E_{br}\sim 1$~TeV, resulting in a $\Gamma_{2,e}\sim 2.3$ higher energy slope. The absence of cooling in the energy band below 1~TeV could be attributed to the escape of the sub-TeV electrons from the system. A similar interpretation of the spectral energy distribution was proposed by~\citet{chernyakova15} for \psrb.

Alternatively, the spectral break in the electron spectrum can occur at the transition between the domination of adiabatic and IC/synchrotron losses 
(see e.g., \citealt{khan07} and \citealt{Suzaku2009} for the \psrb and LS~5039 cases). The adiabatic loss time is naturally shortest in sparse regions outside of the Be star's disk and longest in dense regions inside it.

A broken power-law shape of the spectrum is not unique for the ``similar to \psrb'' model. A similar shape of the spectrum can also be expected within the ``flip-flop'' model, since both interpretations of the break origin can be valid for this model. The two models, however, can be distinguished by CTA observations of the variation of the slope and low-energy break position along the orbit. 

Within the ``flip-flop'' model at orbital phases $\phi=0-0.4$ the compact object moves from a denser to more and more sparser regions of the Be star's disk. The spectrum of relativistic electrons becomes less and less dominated by the losses. This results in a gradual hardening of the TeV slope and a shift of the break energy to higher values. At phases $\phi\sim 0.6-1$ the compact object enters denser regions of the disk which should lead to a gradual softening of the slope/shift of the energy break to lower energies. The spectrum is expected to be the hardest when the object is out of the Be star's disk (orbital phase $\sim 0.4$). Note, that this phase corresponds to the minima of observed emission. The softest spectrum is naturally expected when the compact object approaches the periastron, i.e. at a phase $\phi\sim 0.97$.

In the ``similar to \psrb'' model the compact object intersects the disk of the Be star twice per orbit (at orbital phases $0.2-0.4$ and $0.6-0.8$) where the soft spectrum with the low position of energy break is expected. At phases $0-0.2$, $0.4-0.6$ and $0.8-1$ in the ``similar to \psrb'' model the compact object is out of the dense regions of the disk. At these orbital phases a hard slope with energy break shifted to higher energies can be expected.

\subsubsection{Prospects for CTA observations}

Due to its location, \hess0632 is visible from both the north and south CTA sites (see Table \ref{parameters}). For our simulations we considered two orbital phases: the brightest one ($\phi$=0.2--0.4, hereafter the ``high state'') based on \citet[][]{Aliu2014HESS0632}, and the low-intermediate one ($\phi$=0.8--0.2, hereafter the ``low state'') based on \citet[][]{2017AIPC.1792d0023S}. There are no spectra reported in the literature for the deep minimum state at $\phi$=0.4--0.5, and the two maxima have similar spectra, hence we chose the brightest one as representative of the active state.


\begin{table*}[!ht]
\caption{Best fit of the simulated spectra shown in Fig.~\ref{fig:hess06spe} (0.04-100\,TeV) with a broken power law model for \hess0632. $E_{b}$ (TeV) is the position of energy break, $\Gamma$ is the  photon index  above the break (low energy photon index was frozen to 1.6), F is the 0.04-100\,TeV flux, in $10^{-11}$ ph cm$^{-2}$ s$^{-1}$ units. See text for more details.}\label{J0632::bkpl_sim}
\begin{center}
\begin{small}
	\begin{tabular}{l | ccc |ccc}
	\hline
 \hspace{0.5truecm}  &  \multicolumn{3}{c|}{5h} & \multicolumn{3}{c}{50h} \\ 
Phase	  & $E_{b}$    & $\Gamma$        & Flux    &$E_{b}$ & $\Gamma$           & Flux \\ \hline
	0.2-0.4 & 0.51$\pm$0.10    & 2.30 $\pm$ 0.07 & 1.09 $\pm$ 0.16 & 0.40 $\pm$ 0.04 & 2.30 $\pm$ 0.02 & 1.07 $\pm$ 0.05 \\ 
	0.8-0.2 & 0.46 $\pm$ 0.14  & 2.73 $\pm$ 0.14 & 0.65 $\pm$ 0.15 & 0.40 $\pm$ 0.04 & 2.71 $\pm$ 0.05 & 0.62 $\pm$ 0.04 \\ \hline
	\end{tabular}
\end{small}
\end{center}
\end{table*}


\begin{figure*}[h]
\includegraphics[width=\columnwidth]{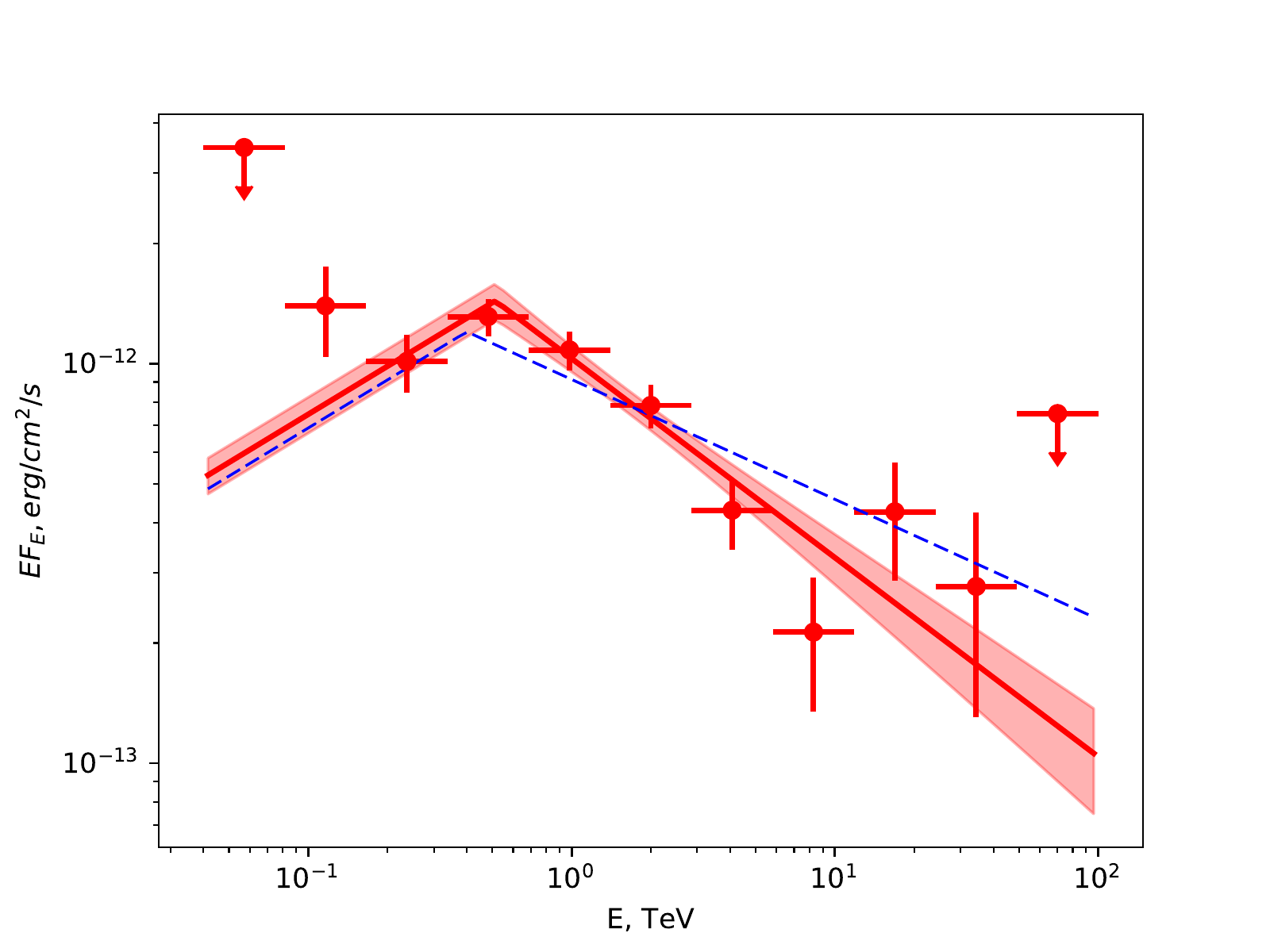}
\includegraphics[width=\columnwidth]{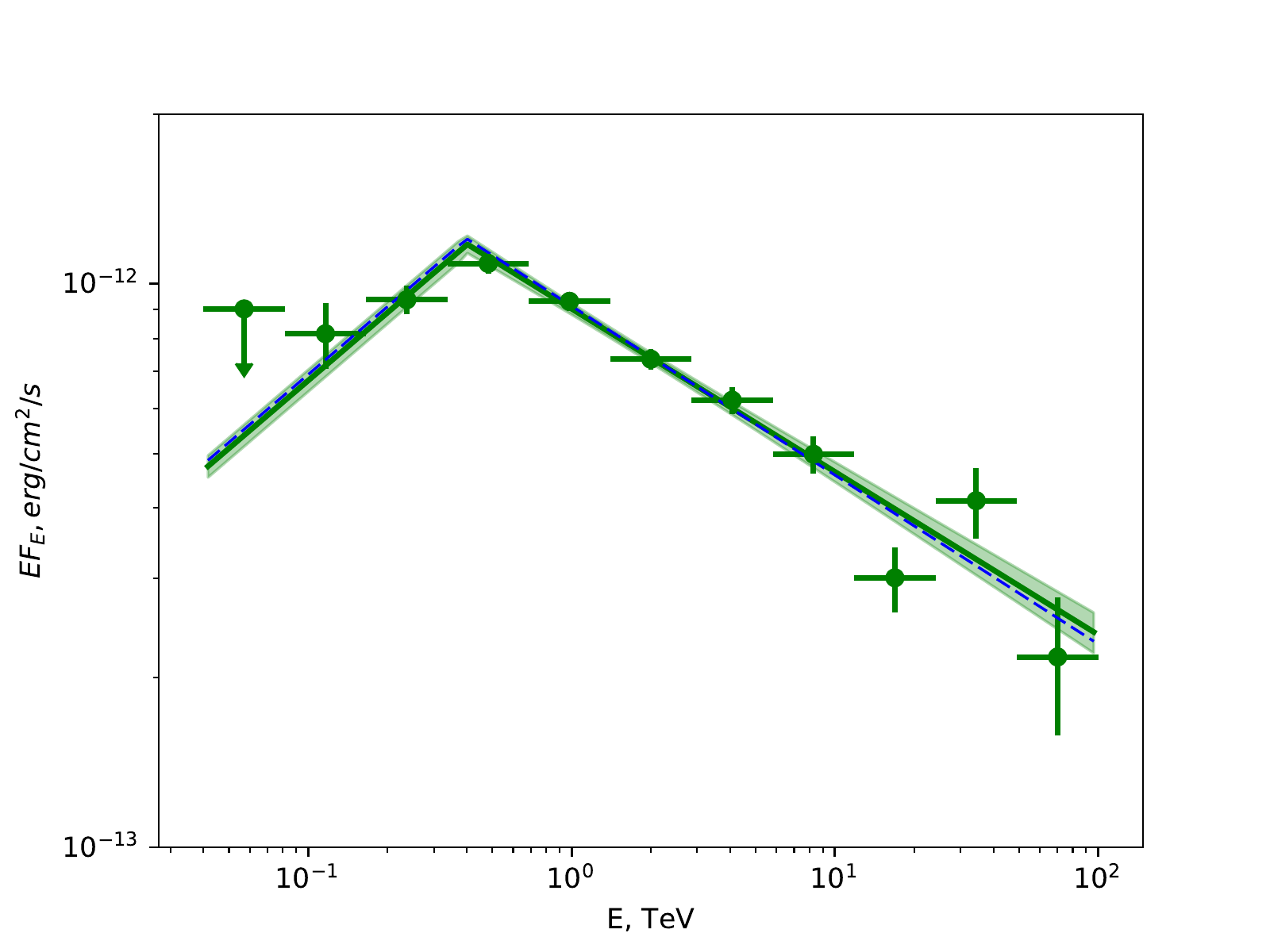}
\includegraphics[width=\columnwidth]{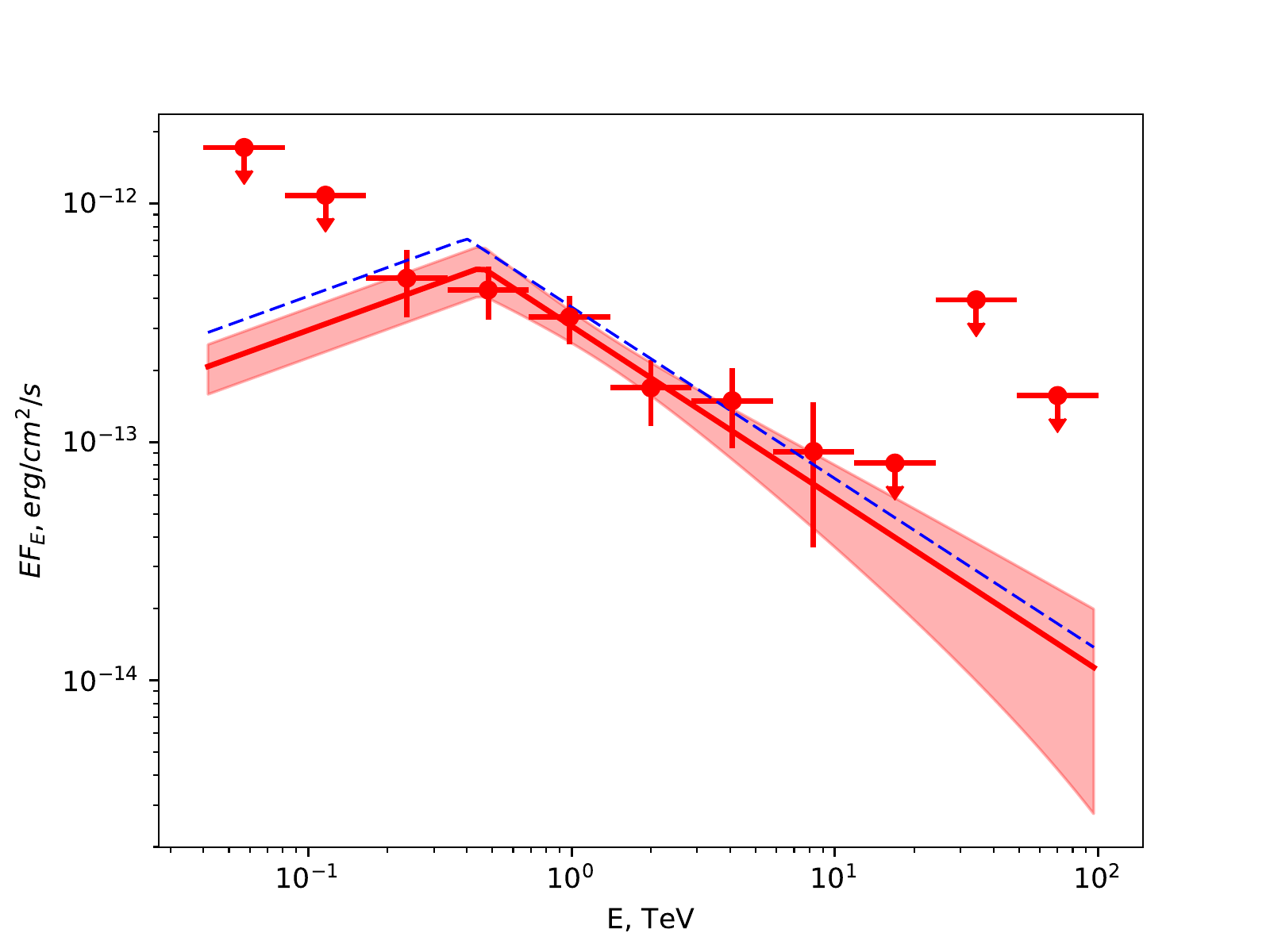}
\includegraphics[width=\columnwidth]{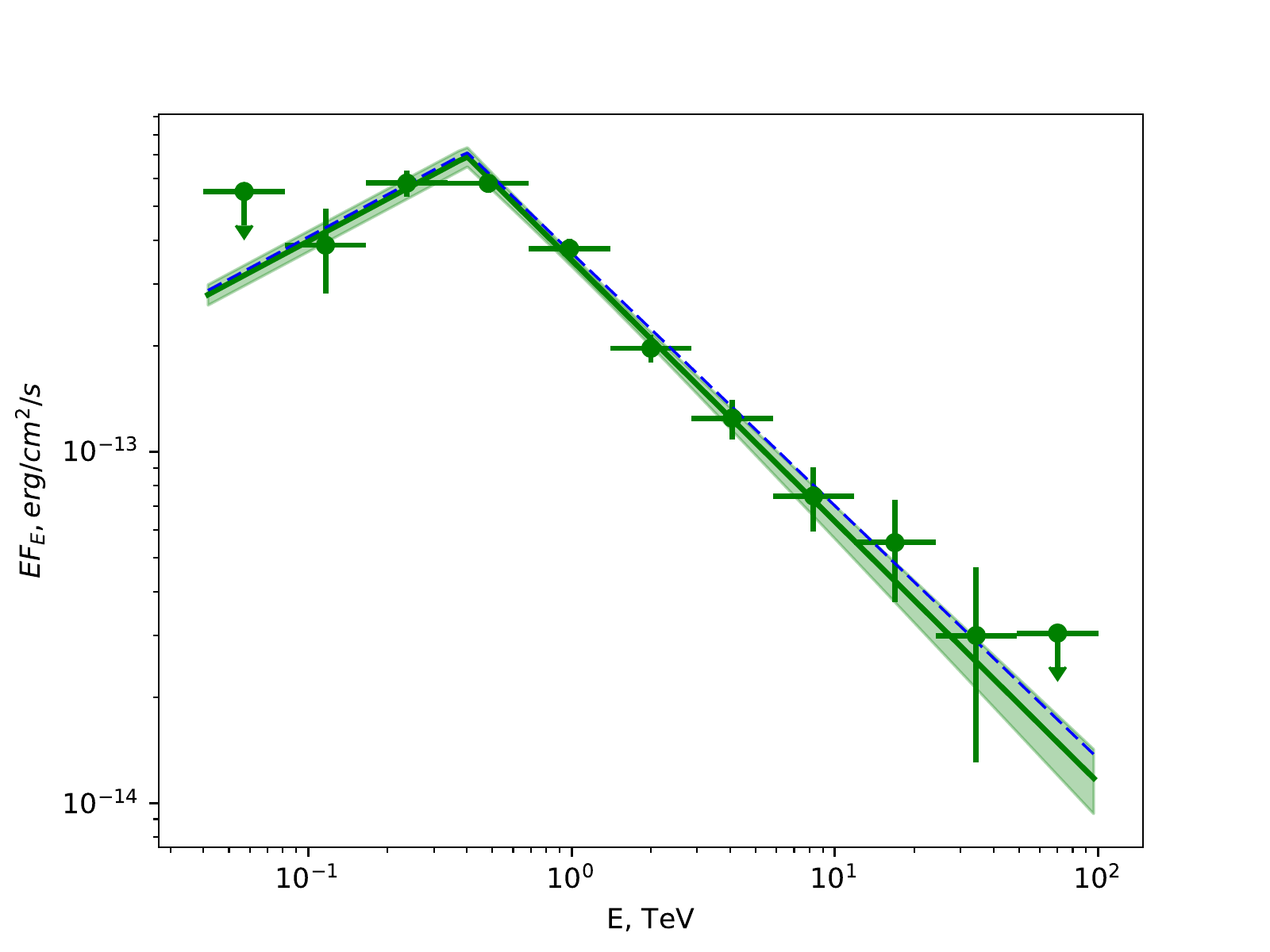}
\caption{Simulated spectra (red and green points) of \hess0632,  {as observed from the Southern site}. In blue the input models are shown. \emph{Upper left:} high state, 5\,h. \emph{Upper right:} high state, 50\,h. \emph{Lower left:} low state, 5\,h. \emph{Lower right:} low state, 50\,h.}\label{fig:hess06spe}
\end{figure*}

In the first group of simulations, we considered the 0.1-100\,TeV energy range.
The source spectral model component was defined as a power-law model with either a photon index 2.3 and normalisation at 1\,TeV of $5.7\times10^{-13}$ ph cm$^{-2}$ s$^{-1}$ TeV$^{-1}$ ($\phi$=0.2--0.4, high state), or a photon index 2.72 and normalisation $2.3\times10^{-13}$ ph cm$^{-2}$ s$^{-1}$ TeV$^{-1}$ ($\phi$=0.8--0.2, low state).

The dependence of the source flux and index uncertainties with different configurations is shown in Fig. \ref{fig:binaries} for 30\,min and 5\,h observations. Simulations also show that a longer 50\,h observation can reconstruct the flux and slope of the source to a better than 3\% accuracy in both the North and South hemispheres.

In the second group of simulations, we simulated the source spectrum with a broken power-law model, in order to study the possibility of the detection of a low energy break. The two physical scenarios discussed above - ``flip-flop'' versus ``similar to PSRB1259-63'' - can be distinguished by CTA observations of the variation of the slope and low-energy break position along the orbit. In both high and low states, we used a low energy slope fixed at 1.6 and an energy break at 0.4\,TeV (the value of the break energy was not fixed in the simulations, so it can vary between different realizations, \citealt{malyshev16}). The high energy slopes are given by the spectral indeces already discussed: $\Gamma$=2.3 for the $\phi$=0.2--0.4 high state, and $\Gamma=2.72$ for the $\phi$=0.8--0.2 low state. The normalizations for the two states at 0.4\,TeV are $4.7\times10^{-12}$ ph cm$^{-2}$ s$^{-1}$ TeV$^{-1}$ (high state), and $2.8\times10^{-12}$ ph cm$^{-2}$ s$^{-1}$ TeV$^{-1}$ (low state). For these simulations, we have focused our attention to the Southern site and the 0.04-100\,TeV energy range. Simulations  of 5\,h and 50\,h observations were performed and the results are shown in Table~\ref{J0632::bkpl_sim}.

Fig.~\ref{fig:hess06spe} shows two single realizations of the spectra of \hess0632, 5\,h (left) versus 50\,h (right). Upper limits are shown when the detection significance is lower than 3$\sigma$. As can be seen, a 50\,h observation will give an excellent reconstruction of the energy break and slopes, whereas a 5\,h observation will suffer from higher uncertainties. 

A direct comparison of the same power-law spectra simulated for CTA with respect to H.E.S.S.\ \citep{Aliu2014HESS0632} and VERITAS \citep{2017AIPC.1792d0023S} observations shows that a CTA snapshot of 5\,h will result in more accurate results than what was previously obtained. Indeed a 55\,h observation of the low state ($\phi$=0.8--0.2) of the source with VERITAS resulted in a $\sim$7\% uncertainty on the detected slope (2.72$\pm$0.2), to be compared to the $\sim$5\% with a 5\,h CTA South observation. 
Similarly, a 15\,h observation of the high state with H.E.S.S.\ resulted in a $\sim$9\% uncertainty on the detected slope (2.3$\pm$0.2), to be compared to the $\sim$3\% with a 5\,h CTA South observation. Note, however, that the CTA error estimates given here are purely statistic, whereas the VERITAS and H.E.S.S. results include systematic errors as well.

As can be seen from Fig. \ref{fig:binaries}
and Table~\ref{J0632::bkpl_sim}, a 5\,h observation will be enough to disentangle the low state from the high state, but it may not be enough to unambiguously disentangle the energy break ($\sim$30\% uncertainty). A 50\,h observation would result in a 10\% uncertainty of the energy break (15\% for a 20\,h observation), allowing the high energy slope and energy break to be accurately monitored along the orbit.  This would enable CTA to disentangle the two currently available scenarios, i.e. ``flip-flop'' versus ``similar to \psrb'' that expect an opposite trend of the spectral slope and energy break from the high state to the low state: a hardening of the spectrum and $E_{b}$ moving to higher energies for the ``similar to \psrb'' scenario versus a softening of the spectrum and $E_{b}$ moving to lower energies in the ``flip-flop'' one.

\hess0632 has a long orbital period ($\sim$316\,d) and each phase will occur only once in one year of observations. Nevertheless, each state is observable for a long period hence a 50\,h observation in the same state (10 nights with $\sim$5\,h each)  is possible. 

\subsection{\hess1832}

\subsubsection{Source properties}

\object{HESS J1832-093}
is a new  $\gamma$-ray binary candidate discovered  as a TeV point source by H.E.S.S. This source lies in the vicinity of SNR G22.7-0.2, which can suggest its possible association with this SNR \citep{abramowski15}. However, several follow-up observations in X-rays  instead support the binary nature of this source~\citep{eger16,mori17}. A simple power law model well describes the TeV  spectrum with a photon index of $\Gamma$ = 2.6 $\pm$ 0.3stat $\pm$ 0.1sys and an integrated photon flux above 1 TeV of $F = (3.0 \pm 0.8\mbox{stat} \pm 0.6\mbox{syst} )\times 10^{-13}$~cm$^{-2}$~s$^{-1}$~\citep{abramowski15}. An {\it XMM-Newton} observation of the source field discovered a bright X-ray source, XMMU J183245-0921539 within the $\gamma$-ray error circle~\citep{abramowski15}. This source is also associated with a point source detected in a subsequent \cha observation campaign~\citep{eger16}. During the \cha observations, an increase of the 2-10 keV flux of the order of 4 with respect to the earlier \xmm measurement and the coincidence of a bright IR source at the \cha error box suggest a binary scenario for the $\gamma$-ray emission~\citep{eger16}.  

Recently \citet{mori17} reported on a \nus X-ray observation of the field containing \hess1832 and a re-analysis of the archival \cha and \xmm  data.  The data reanalysis does not confirm the same level of flux variation reported in previous work. However \citet{mori17} found other evidences supporting the $\gamma$-ray binary scenario for this source: the X-ray \nus spectrum extends up to 30 keV and the best fit is represented by a simple power law model with a photon index of $\Gamma$=1.5 without any break or cut-off. The \nus 2-10 keV flux is 1.5 times higher than the 2011 \xmm flux in the same range. Even if no pulsations were detected in the power spectrum, the flat power density spectrum underlines the lack of accretion-powered emission from the source. The authors conclude that the X-ray timing properties of \hess1832 are similar to those of the other $\gamma$-ray binaries.
Note however, that at the moment no GeV emission has been detected from the system 
~\citep{eger16,abramowski15}.  A sure identification of \hess1832 as a $\gamma$-ray binary still needs the detection of the orbital period as well as simultaneous X-ray and $\gamma$-ray observations.

During the H.E.S.S.\ measurement no significant flux variability was detected in the long term light curves \citep[run, day, month, as reported by][]{2015A&A...577A.131H}. The data set consisted of 67 h of observations taken from 2004 to 2011. Even if the data sample has a low density, the lack of flux variability could imply that the eventual modulation of the $\gamma$-ray flux is probably quite smooth and mostly within the H.E.S.S.\  flux error (the H.E.S.S.\ statistical error is about 26\%). Even if the X and $\gamma$-ray characteristics resemble those of HESS J0632+057, the orbital period should be long, similar to e.g. \psrb.

\subsubsection{Prospects for CTA observations}

We performed two sets of a 1000 simulations for 0.5~h and 5\,h exposure time respectively.
The input source spectrum has a power law shape with a slope $\Gamma$=2.6,
as reported by \citet{2015A&A...577A.131H}.  
The simulation results are shown in Fig.~\ref{fig:binaries}. The results of our simulations show that the source is too faint to be firmly detected in 30 min, but for 5 h we obtained a $\sim$14$\sigma$ detection. Moreover CTA observations of \hess1832 will surely improve the angular resolution allowing further constraints on the extension of the TeV source. The spectra measurement will be extended both below and above the H.E.S.S.\ detection, which will allow eventual detection of a spectral cut-off. Simulations reported in Fig. \ref{fig:binaries} demonstrate that with a 5 h exposure CTA will be able to detect a high energy cut-off if it is present below 10 TeV. Finally, the high CTA sensitivity could detect flux variation and spectral modulation. In particular, a detection of flux modulation will allow us to fix the orbital period, that is the first step for a sure identification of \hess1832 with a $\gamma$-ray binary.

\subsection{\lmc}

\subsubsection{Source properties}

\begin{figure*}[h]
\includegraphics[width=\columnwidth]{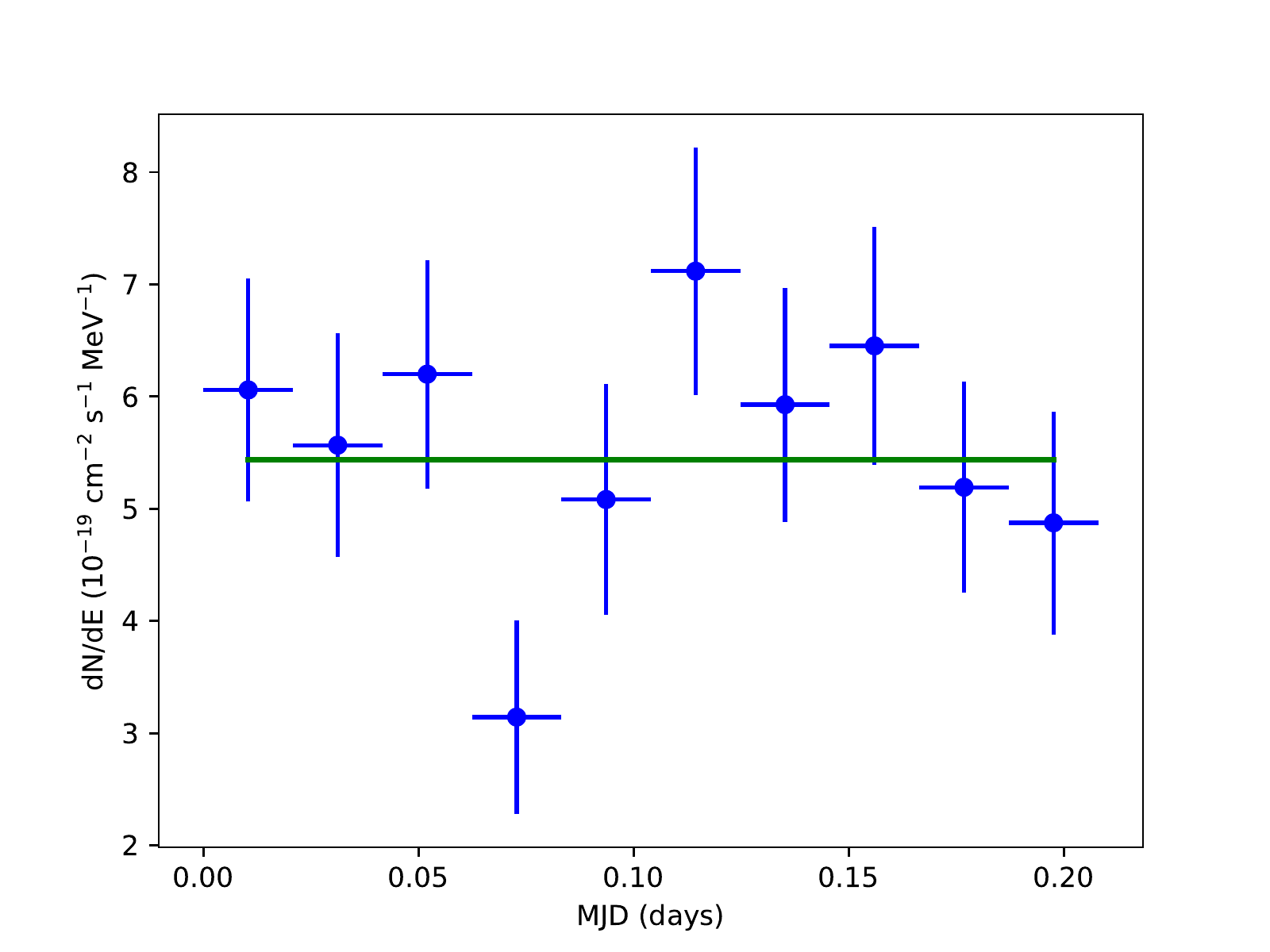}
\includegraphics[width=\columnwidth]{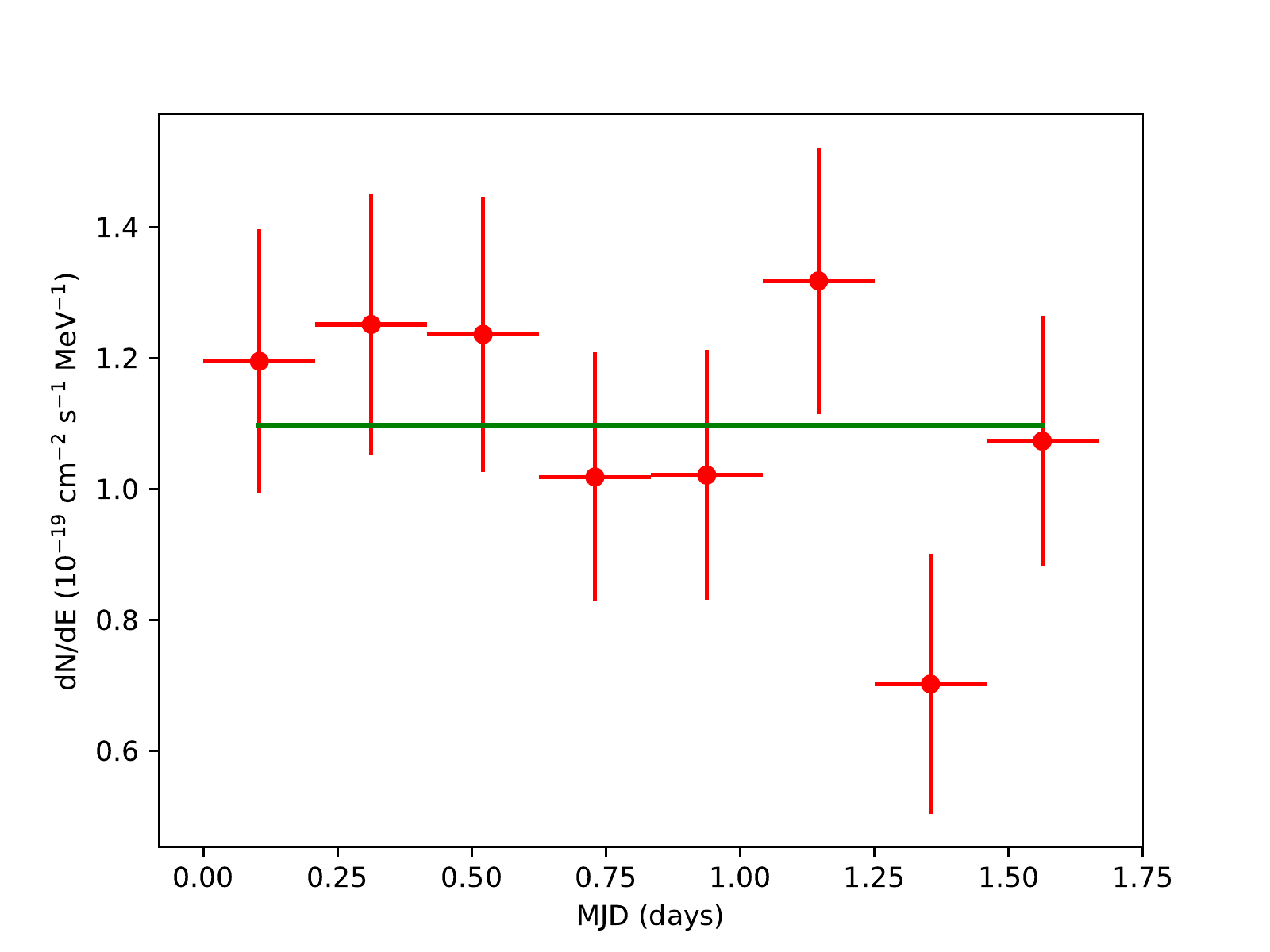}
\caption{Simulated CTA light curves of \lmc at high (left panel, 30 min time bin) and low (right panel, 5 h time bin) states. }\label{fig:lmcp3:lc}
\end{figure*}

\object{LMC P3} is the first and, up to the moment, the only known extragalactic \grb. It was detected in 2016 with the \flat in the Large Magellanic Cloud from a search for periodic modulation in all sources in the third \flat catalog \citep{3FGL}. The system has an orbital period of 10.3 days and is associated with a massive O5III star located in the supernova remnant DEM L241 \citep{2016ApJ...829..105C}. \swift/XRT X-ray  and ATCA radio observations demonstrated that both X-ray and radio emission are also modulated on the 10.3 day period, but are in anti-phase with the $\gamma$-ray modulation. The X-ray spectrum is well described by a single power law with $\Gamma=1.3\pm0.3$, modified by a fully covered absorber. The resulting value of the hydrogen column density of a fully covered absorber is comparable with the Galactic HI value.

Optical radial velocity measurements suggest that, unless the system has a very low inclination the system contains a neutron star \citep{2016ApJ...829..105C}. Low inclinations, however, result in a range  of masses of the compact object above the Chandrasekhar limit, e.g. a BH with a mass of  M=5M$_\odot$ will have an inclination $i=14{^{+4}_{-3}}^\circ$, and   $i=8\pm 2^\circ$  for M=10M$_\odot$.  The source is significantly more luminous than similar sources in the Milky Way at radio, optical, X-ray and $\gamma$-ray wavelengths. It is at least four times more luminous in GeV gamma rays and 10 times more luminous in radio and X-rays than LS 5039 and 1FGL~J1018.6-5856, though the luminosity of the companion star and the orbital separations are comparable in all three systems.

The LMC has been observed extensively with H.E.S.S. since 2004. The data which were collected for the LMC between 2004 and the beginning of 2016 results in an effective exposure time for LMC P3 of 100 hours \citep{2018arXiv180106322H}. The sensitivity of H.E.S.S. does not allow a detection of flux variations of the object on a nightly basis. 
The low flux coming from the system does not allow for any statistically significant detection of periodicity using a Lomb-Scargle test and the Z-Transformed Discrete Correlation Function.
Folding the light curve with the orbital period of the system of 10.301 days, clearly demonstrates the orbital modulation of the VHE with a significant detection only in the orbital phase bin between 0.2 and 0.4 (orbital phase zero is defined as the maximum of the HE light curve at MJD 57410.25). The H.E.S.S. spectrum during the on-peak part of the orbit is described by a power-law with a photon index $\Gamma=2.1\pm0.2$. The averaged slope along the total orbit is softer with $\Gamma=2.5\pm0.2$. The VHE flux above 1 TeV varies by a factor more than 5 between on-peak and off-peak parts of the orbit.

The minimum HE emission occurs between orbital phases 0.3 - 0.7. 
The shift between the orbital phase of HE and VHE peaks is not unique to this $\gamma$-ray binary.
For example, 
a similar shift
is observed in \ls (see Section \ref{section:ls}), as the angle-dependent cross section of IC scattering and $\gamma \gamma$ absorption due to pair-production affects the HE and VHE in  different ways~\citep[e.g.][]{LS5039_2008_Dubus,LS5039_2008_Khan,2008ApJ...672L.123N}.

Recently reported optical spectroscopic observations of \lmc\ have better constrained the orbital parameters \citep{2019MNRAS.484.4347V}. The observations find the binary has an eccentricity of $0.4\pm0.07$ and place superior conjunction at phase $\sim0.98$ and inferior conjunction at phase $\sim 0.24$. These phases correspond to the points of the maxima reported in \flat and H.E.S.S.\ light curves respectively. The mass function found ($\sim 0.0010 $ M$_\odot$) favours a neutron star companion, for most inclination angles.

The detection of VHE emission during the entire orbit is critical for detailed modelling that will allow us to understand what is happening in the system.

\subsubsection{Prospects for CTA observations}
Following the results of H.E.S.S. observations, we have simulated light curves and spectra that CTA will observe during the high ($F_{\rm TeV}=5\times 10^{-13}$ cm$^{-2}$s$^{-1}$) and low ($F_{\rm TeV}=1\times 10^{-13}$ cm$^{-2}$s$^{-1}$) states, where $F_{\rm TeV}$ is the source flux above 1TeV. In our simulations, we assumed a constant flux within each of the states. The results are presented in Fig. \ref{fig:lmcp3:lc}. During the high state CTA will be able to detect variability of the source at a  3 $\sigma$ confidence level if it is higher than 60$\%$ on a 30 min time scale, and  40$\%$ on a 1 hour time scale.

\begin{figure}[h]
\includegraphics[width=\columnwidth]{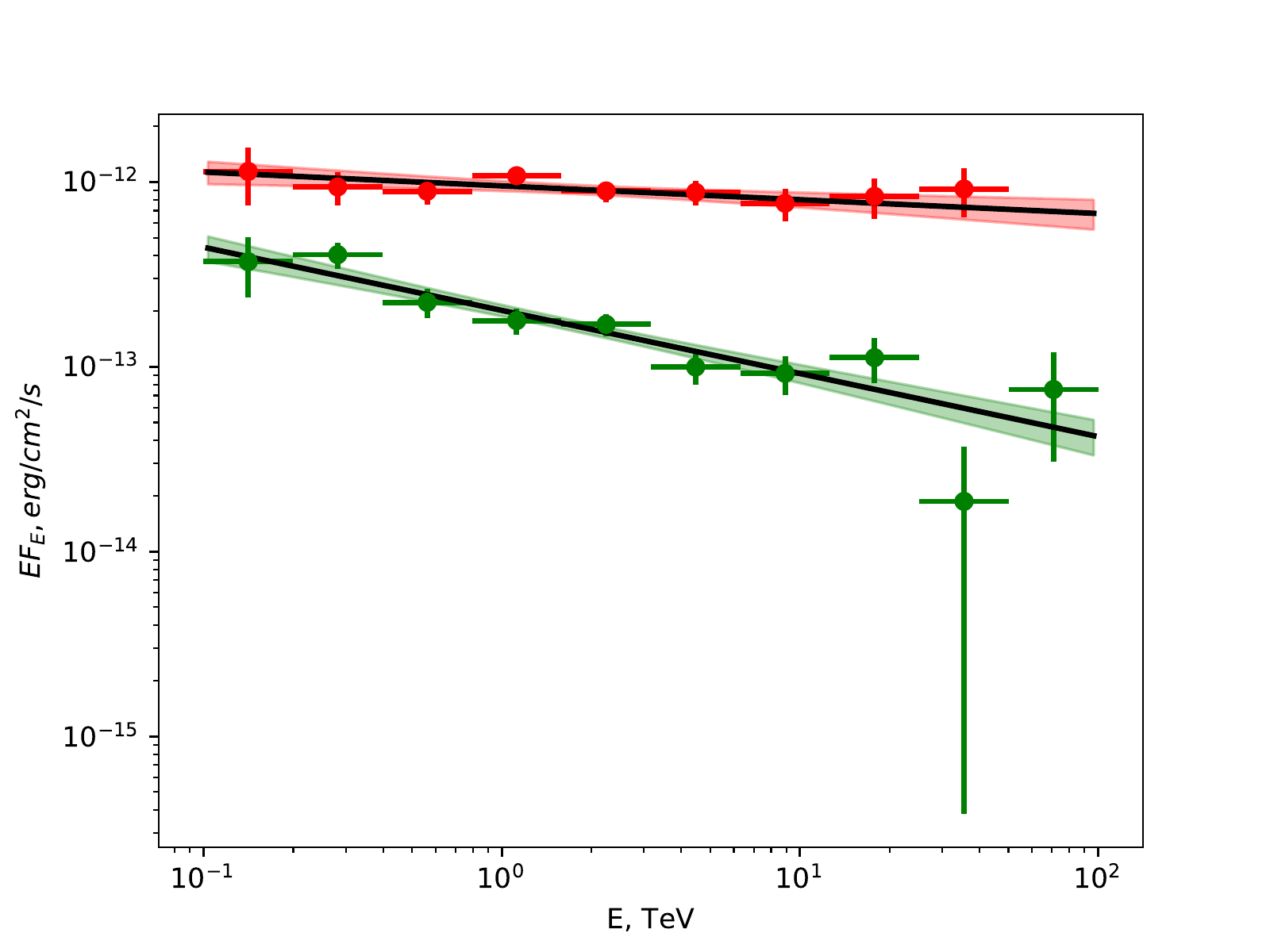}
\caption{Simulated CTA spectrum of \lmc at high (red points, 5 h exposure) and low (green points, 40 hours exposure) states. }\label{fig:lmcp3:spec}
\end{figure}

In Fig. \ref{fig:lmcp3:spec} we show the spectra of the high and low states of the source. We will be able to determine the slope of the spectrum with an accuracy of 2$\%$  for 5 h of observations (6$\%$ for 1 hour) during the high state, and 2$\%$ for 40 hours observations (10$\%$ for 5 h) during the low state.

Thus the CTA sensitivity will be high enough to study the nightly averaged spectral evolution of the source with an accuracy of better than 10$\%$ along the orbit. This will allow us to understand the details of the physical processes in this system and develop a consistent model of the multi wavelength emission. 

\section{Colliding wind binaries (CWBs) \label{sec::cwb}}
\subsection{\etacar \label{etacar}}

\object{$\eta$~Carinae} 
is the most luminous massive binary system in our Galaxy,  
and the first binary which does not host a compact object that has been detected at very high energies.
It is believed to be composed of a Luminous Blue Variable (LBV), possibly originating as a star with an initial mass $\gtrsim 90 M_{\odot}$ \citep{2001ApJ...553..837H}, and of a Wolf-Rayet (WR) companion. The former is accelerating a very dense wind with a mass loss rate of $\sim 
8.5\times 10^{-4}$ M$_{\odot}$ yr$^{-1}$ and a 
terminal velocity of $\sim 420$ km s$^{-1}$ \citep{2012MNRAS.423.1623G}. Based on the observed X-ray emission of this system, it is believed that its companion also emits a powerful wind, with a mass loss rates of $\simeq 10^{-5}$ M$_{\odot}$ yr$^{-1}$ at a velocity of 3000 km s$^{-1}$ \citep[][see Table \ref{parameters-cwb}]{2002A&A...383..636P,2005ApJ...624..973V,2005MNRAS.357..895F,2009MNRAS.394.1758P}.

\begin{table*}
\caption{System parameters of colliding wind binaries\label{parameters-cwb} }
\label{table:parameters-cwb}
\begin{center}
\begin{tabular}{@{}llll@{}}
\hline
                              & $\eta$ Carinae$^+$ & $\gamma^2$ Vel$^{\star}$  & HD 93129A$^{\diamond}$\\
 \hline
 P$_{\rm orb}$ (days)         &	$2022.7\pm1.3$  & $78.53 \pm 0.01$     & $121^{+60}_{-39}$ yr      \\
 $T_{0}$                   & 2020.11      	& $50120.4 \pm 0.4$ MJD & $2017.60^{+0.38}_{-0.32}$  \\
 $e$ 				          &    	0.9 - 0.95  & $0.334 \pm 0.003$     & $0.967^{+0.023}_{-0.026}$  \\
 $\omega$ (\degr)	    &  240 -- 285  & $67.4\pm 0.5$      & $177^{+153}_{-143}$      \\
 $i$ (\degr)		          &     130-145     & $65.5\pm 0.4$     &  $117^{+28}_{-7}$   \\
 $d$ (kpc)			          & 	2.3         &  $0.336^{+0.008}_{-0.007}$    & $3.093^{+0.335}_{-0.276}$$^{**}$  \\
 \hline
   Star A		              & eta Car A	   & WR11	     &  HD 93129Aa \\
spectral type 		          &	  LBV          & WC8 	     &   O2 If*\\
$M_{\star}$ (M$_{\odot}$)     & 	90         & $9.0\pm 0.6$ 	 &  60-110	\\
$R_{\star}$ (R$_{\odot}$)     &  	60-100     & $6\pm 3$        &   18.3\\
$L_{\star}$ ($10^5L_{\odot}$) &      50        & 1.7	     &   25\\
$\dot M$ $(10^{-6}M_{\odot}yr^{-1}$) & 250-1000 & $8\pm 4$   	 &   10 \\
$V_{\infty}$ (km/s)           &         500    &$1550\pm 150$    &   3200\\
 \hline
 Star B 		              & eta Car B      & 	         &   HD 93129Ab \\
 spectral type 		          &	   (WR/O)      & O7.5	     &   O3.5 V\\
 $M_{\star}$ (M$_{\odot}$)    & 	   30      & $28.5\pm 1.1$   &   30-70 \\
$R_{\star}$ (R$_{\odot}$)     & 	14.3–23.6  & $17\pm 2$  	 &   13.1 \\
$L_{\star}$ ($10^5L_{\odot}$) &         10     & 2.8         &   55 \\
$\dot M$ $(10^{-6}M_{\odot}yr^{-1}$) & 10-15   &0.18 $\pm$ 0.04         &   5.3  \\
$V_{\infty}$ (km/s)           &       3000     &2500 $\pm$ 250    &   3000 \\
 \hline
$d_{\rm periastron}$ (R$_{\odot}$)     &   331   &  172	 &  870 \\
$d_{\rm apastron}$ (R$_{\odot}$)       &   3642   &  344	 &      \\
\hline
 IRF: 	                      &   South\_z40    & South\_z40  &   South\_z40   \\     
\hline
\multicolumn{4}{l}{ $^+$~See references in the text.}\\
  \multicolumn{4}{l}{ $\star$ \citet{1997A&A...328..219S,1999A&A...345..163D, 2007MNRAS.377..415N}.}\\
\multicolumn{4}{l}{ $^{\diamond}$ \citet{2016A&A...591A.139D,2017AIPC.1792d0027D,2017MNRAS.464.3561M}.}\\
\multicolumn{4}{l}{$^{**}$ The Gaia archive, https://gea.esac.esa.int/archive/}

\end{tabular}
\end{center}
\end{table*}

The inference of binarity for $\eta$ Car is indirect, given that the companion is not  directly observed at any wavelength. Spectral periodic variations have been used to indicate its binary nature, and to provide estimates of both the orbital and the companion stellar parameters. The  modulation detected in the X-ray light curve indicates that the two stars are on a very eccentric orbit \citep{2001ApJ...547.1034C,2005MNRAS.357..895F,2008MNRAS.388L..39O}, while its hardness and intensity allow for the estimates of the wind velocity and mass-loss rate shown in Table \ref{table:parameters-cwb}.

 During its Great Eruption (1837-1856), $\eta$ Car ejected $10-40~M_\odot$ \citep{2010MNRAS.401L..48G} at an average speed of $\sim 650$ km s$^{-1}$ \citep{2003AJ....125.1458S}, forming the Homunculus Nebula and releasing $10^{49-50}$ erg of energy. The  orbital period at the epoch of the Great Eruption was $\sim$~5.1 yr, and increased up to the current $\sim~5.54$~yr \citep{2004MNRAS.352..447W,2005AJ....129.2018C,2008MNRAS.384.1649D}. Two other major eruptions occurred since then \citep{2014ApJ...791...95A}, resulting in ejecta that interact with the radiation coming from within. The long-term X-ray modulation observed for different orbits is possibly due to the time evolution of the ejecta.  

The relative separation of the two stars varies by a factor $\sim10 - 20$, depending on the estimated eccentricity ($e \sim 0.9 - 0.95$). At periastron, the two objects pass within a few AU of each other, a distance just a few times larger than the size of the primary star. In these extreme conditions their supersonic winds  form a colliding wind region of hot shocked gas where charged particles can be accelerated via diffusive shock acceleration up to high energies \citep{eichler93,2003A&A...409..217D,2006ApJ...644.1118R}. As these particles encounter conditions that vary along the orbit, one can expect an orbital dependency of the $\gamma$-ray emission.

The hard X-ray emission detected by INTEGRAL \citep{2008A&A...477L..29L} and Suzaku \citep{2008MNRAS.388L..39O}, with an average luminosity $(4$-$7)\times10^{33}$ erg s$^{-1}$, suggested the presence of relativistic particles in the system. AGILE detected a variable $\gamma$-ray source compatible with the position of $\eta$ Car \citep{2009ApJ...698L.142T}. \flat detected very energetic emission $\gtrsim 10$ GeV around periastron \citep{2010ApJ...723..649A,2011A&A...526A..57F,2012A&A...544A..98R}, that can be interpreted as the $\pi^0$-decay of accelerated hadrons interacting with the dense stellar wind \citep{2011A&A...526A..57F}. Other authors assume that the intrinsic cutoff of the $\gamma$-ray spectrum can be placed at higher energies ($250\sim500~\mathrm{GeV}$), whilst the observed final spectrum is the consequence of the $\gamma-\gamma$ absorption over an ad hoc distribution of soft X-ray photons \citep{2012A&A...544A..98R}.

Cherenkov observations \citep{2012MNRAS.424..128H,2017arXiv170801033L} imply a sudden drop in the spectrum at energies $\gtrsim 1$ TeV, that could be interpreted as a cut-off in the accelerated particle distribution or due to severe $\gamma-\gamma$ absorption. 

\subsubsection*{$\gamma$-ray Variability}

\cite{2011ApJ...726..105P} presented three dimensional hydrodynamical simulations of $\eta$ Car 
which reproduced the 
observed X-ray spectra and light-curves. The acceleration of particles to relativistic energies depends also on the magnetic field of the shock region. The first magneto-hydrodynamical simulations of the colliding winds in $\eta$ Car was performed by \cite{2012MNRAS.423.1562F}. The authors showed that the amplification factor of the field, within the shocks, is orders of magnitude larger than the estimates from Rankine-Hugoniot jump conditions because of the pile-up effect \citep{2015MNRAS.446..104R}. 

With respect to the energetic particles, \cite{2015wrs..conf..289F} numerically integrated particle trajectories on top of the previous MHD simulations in order to study their acceleration. The author showed that the complex geometry of the field at the cooled shock region results in diffuse acceleration to be more efficient compared to the first order Fermi process. Maximum energies of $1-10$~TeV where obtained for the typical wind parameters of the system. Unfortunately the number of particles simulated was insufficient to fully predict the probability distribution function of the energetic particles, as well as the consequent spectral energy distribution of radiation. To estimate the non-thermal emission of the system, \cite{2017A&A...603A.111B} calculated the maximum energies reached by electrons and hadrons cell-by-cell assuming a dipolar magnetic field at the surface of the main star. The magnetic field is the only additional parameter and can be tuned. Shock velocities and mechanical power were calculated in every cell, including those outside the shock region. Most of the shock power is released on both sides of the wind collision zone and in the cells downstream of the wind-collision region \citep{2006ApJ...644.1118R}. The increasing shock area compensates for the loss of the released energy density up to a relatively large distance from the centre of mass, explaining why the X-ray luminosity at apastron is still about a third of the peak emission at periastron. 

The optical depth of the wind for $\gamma$-ray absorption varies between 10$^{-6}$ at apastron and $\sim$10$^{-2}$ at periastron, excluding that the 1-100 GeV spectral shape could be explained by absorption \citep{2012A&A...544A..98R}.

The mechanical luminosity available in the shock increases towards periastron (the same trend is followed by the thermal emission) and almost doubles in the phase range $\approx 1.05 - 1.15$. The latter peak corresponds to a bubble with reverse wind conditions developing and effectively doubling the shock front area during about a tenth of the orbit \citep{2011ApJ...726..105P}. The density of this bubble is low, its thermal emission does not contribute significantly. The mechanical luminosity shows a local minimum between phases 1.0 and 1.05, when the central part of the wind collision zone is disrupted. Note that phases 1 and 2 correspond to first (2009) and second (2014) periastron observed during \flat operation epoch, respectively.

\begin{figure}[b]
\includegraphics[width=\columnwidth]{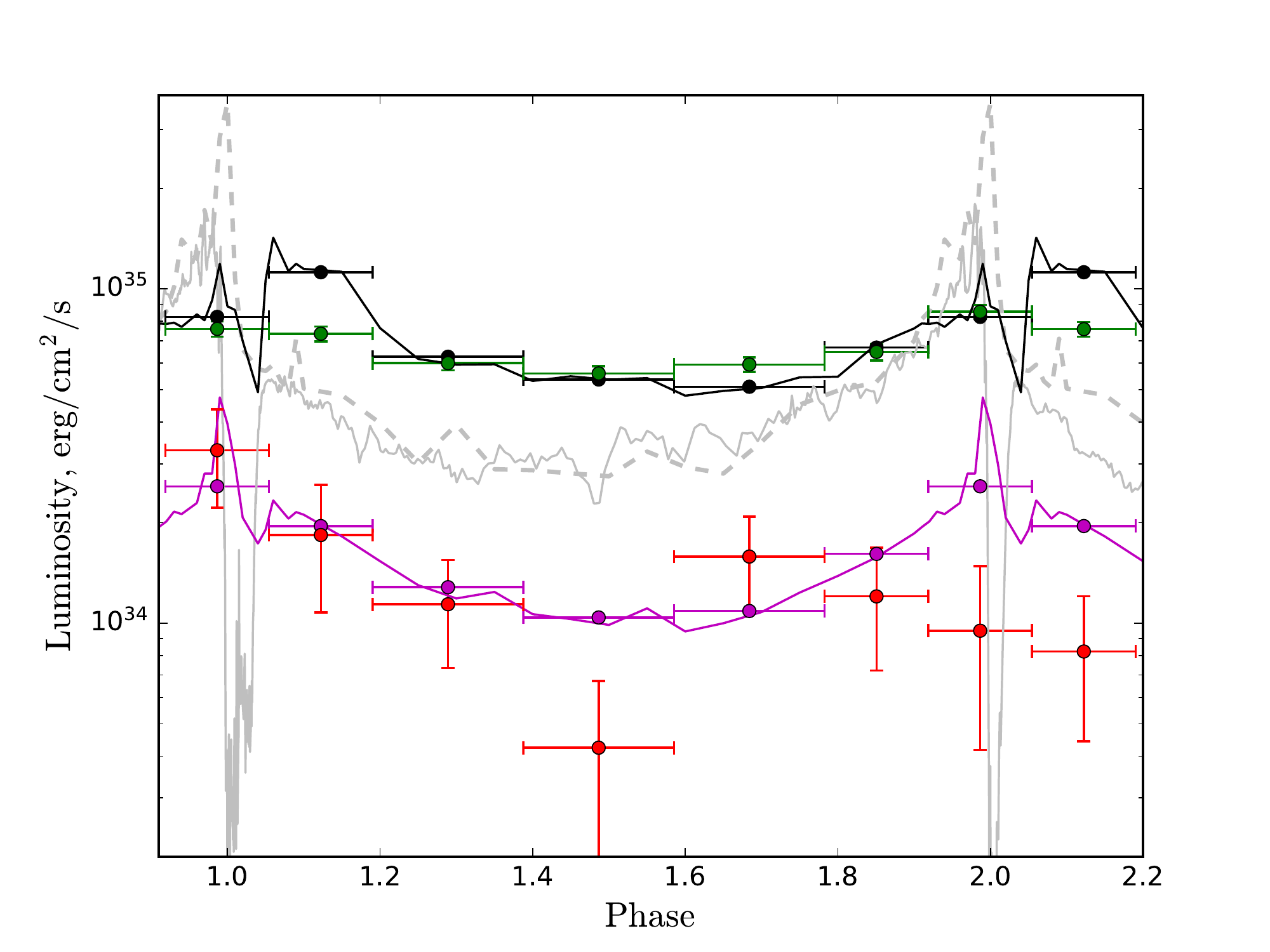}
\caption{Simulated and observed X-ray and $\gamma$-ray light-curves of $\eta$ Car. The black and purple lines and bins show the predicted inverse-Compton and neutral pion decay light-curves. The green and red points show the observed \flat light-curves at low (0.3-10 GeV) and high (10-300 GeV) energies. The dim grey light-curves show the observed (continuous) and predicted (dash, without obscuration) thermal X-ray light-curves. Error bars are $1\sigma$. Phases 1 and 2 correspond to first (2009) and second (2014) periastron observed during \flat operation epoch, respectively \citep[from][]{2017A&A...603A.111B}.}\label{fig:simul}
\end{figure}

Electron cooling, through inverse-Compton scattering, is very efficient and the induced $\gamma$-ray flux peaks just before periastron. Because of the bubble mentioned above a secondary inverse-Compton peak is expected above phase 1.05 although its spectral shape could be different. The relative importance of the secondary peak depends on the magnetic field geometry, radiation transfer, obscuration and details of the hydrodynamics. The situation is different for hadrons. Unless the magnetic field 
is 
very strong ($>$ kG) hadronic interactions mostly take place close to the centre and a single peak of neutral pion decay is expected before periastron.

Fig. \ref{fig:simul} shows the X-ray and $\gamma$-ray light curves predicted by the simulations for a surface magnetic field of 500 G and assuming that 1.5\% and 2.4\% of the mechanical energy is used to respectively accelerate electrons and protons. To ease the comparison between observations and simulations, the results of the latter were binned in the same way as the observed data.

Such a surface magnetic field provides a good match to the observations but magnetic field amplification at the shock could scale it down. The predicted flux at phase 1.1 is twice too large when compared with the observation. This discrepancy largely comes from the energy released in the inverted wind bubble after periastron. The ratio of the emission generated in the shocks on both sides of the wind collision zone is relatively constant along the orbit except at phase 1.1, where much more power is generated in the shock occurring in the wind of the secondary star. This discrepancy may indicate that the inverted bubble is either unstable or produces a significantly different inverse-Compton spectrum.

Observations match the predictions of the simulation except for the second peak, which is slightly shifted towards earlier phases and has a lower luminosity \citep[see Fig. 6 in ][]{2017A&A...603A.111B}.
The phase difference could be related to the eccentricity $(e=0.9)$ assumed in the simulation, which is not well constrained
by observations
\citep{2000ApJ...528L.101D,2001ApJ...547.1034C}, and that has an important effect on the inner shock geometry. 

The distribution of the maximum electron energy, weighted by the inverse Compton emissivity, and hence the resulting photon distribution, are quite smooth. The difference in the electron spectral shape on both sides of the wind collision zone cannot account for the two $\gamma$-ray components as suggested by \cite{2011A&A...530A..49B}, who assumed a simplified geometry.

The inverse-Compton emission peaks slightly below 1 GeV and does not extend beyond 10 GeV at the level observed during the first periastron \citep[see Fig. 4 in][]{2018arXiv181004168W}, contrasting with the conclusions from \cite{2015MNRAS.449L.132O}, 
who attributed 
the full \flat detection to hadronic emission. Their simulations predict a smaller variation between periastron and apastron, a longer flare around periastron and a deeper minimum when compared to the observed data. Such discrepancies might be due to the simplified geometry assumed by the authors and by the artificially reduced particle acceleration at periastron. Inverse-Compton emission and neutral pion decay \citep{2011A&A...526A..57F} remain, therefore, a good model of the $\gamma$-ray variability.

The simulated pion induced $\gamma$-ray light-curve and its variability amplitude show a single peak of emission centred at periastron, in good agreement with the \flat observations of the first periastron. These simulations predict that the hadronic cut-off energy varies between 200 TeV and 2 TeV from periastron to apastron. $\eta$ Car may therefore accelerate particles close to the knee of the cosmic-ray spectrum. 

The second periastron is different, with a lack of high energy emission. It has been suggested that the change 
in the X-ray emission after that periastron (a significant decrease can be observed in Fig. \ref{fig:simul}, see also \citealt{2015arXiv150707961C}) was the signature of a change of the wind geometry, possibly because of cooling instabilities. A stronger disruption or clumpier wind after the second periastron could perhaps induce a decrease of the average wind density and explain 
why less hadronic interactions and less thermal emission took place, without 
greatly affecting   
the inverse-Compton emission.

\subsubsection{Prospects for CTA observations}

\begin{figure}[t]
\includegraphics[width=\columnwidth]{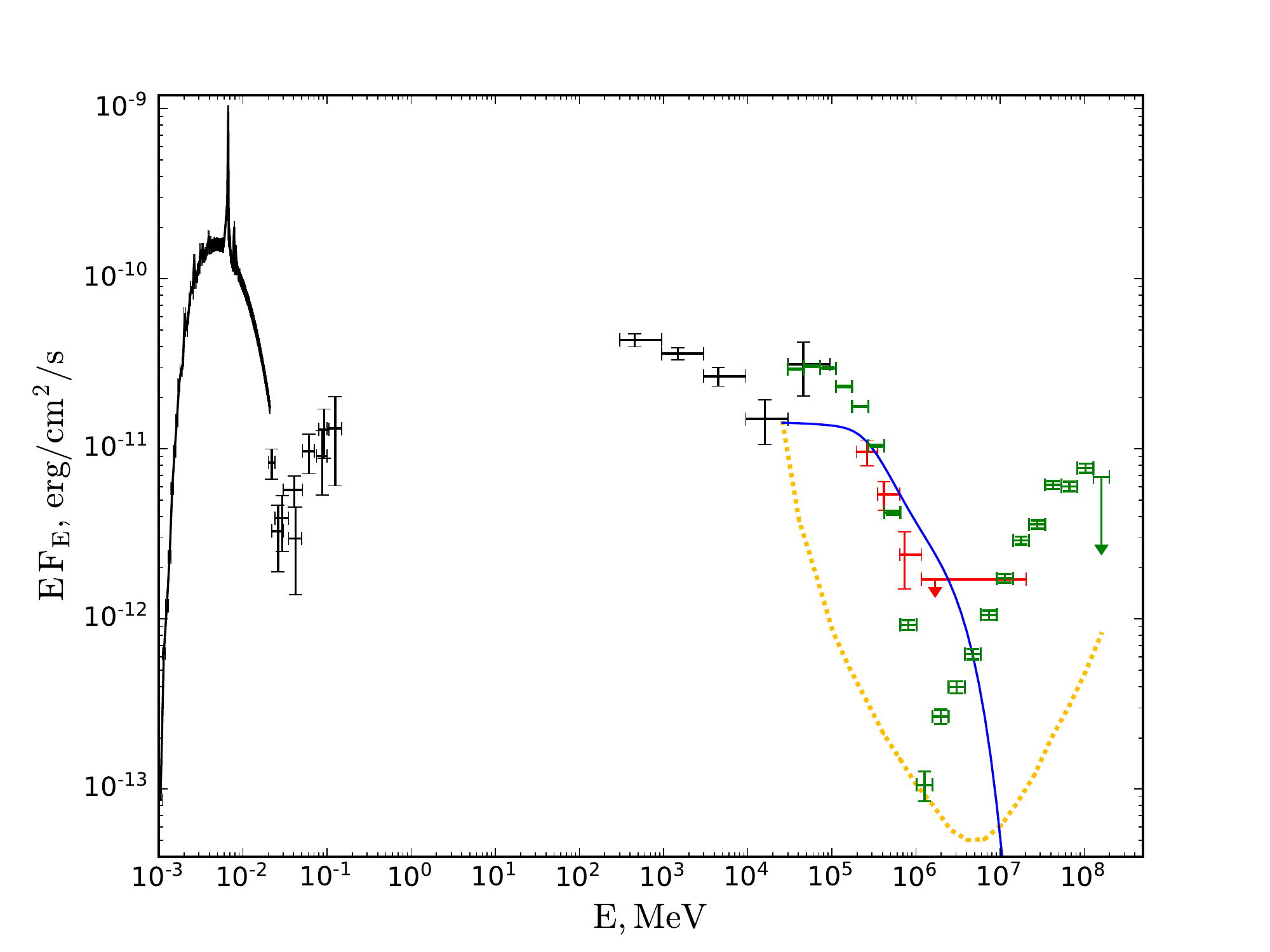}
\caption{Shock induced spectral energy distribution in $\eta$ Car from 1 keV to 1 PeV. The data (black) are from NuStar \citep{2018A&A...610A..37P}, Swift/BAT, INTEGRAL and \flat. The red points are the H.E.S.S. measurements obtained close to periastron. Green points show a 50 hours simulated observation by CTA at periastron. The blue line shows the spectrum which could be expected at apastron with a lower energy cut off and obscuration. The system geometry can be constrained by a shift of the obscuration peak along the orbit, which is not taken into account here. The yellow dotted line is the 50h CTA point source sensitivity ($5\sigma$ per bin of $\Delta E/E=0.2$).}
\label{fig:sed}
\end{figure}

The CTA telescopes to be placed in the southern hemisphere will be very sensitive to probe the spectrum and variability of $\eta$ Car above 30 GeV. As discussed above the contribution of the IC emission is negligible above 10 GeV and the observed VHE emission is only due to $\pi^0$ decay. The $\pi^0$ decay spectrum should be heavily modified by $\gamma$-$\gamma$ absorption of TeV photons against the numerous ultraviolet photons emitted by the stellar surfaces. The photon field is probably opaque at 1 TeV with an optical depth varying quickly depending on the relative position of the two stars and of the line of sight. The intrinsic $\pi^0$ decay spectrum is a complex convolution of the maximum energy, luminosity, particle drift and obscuration.  
The obscuration is likely maximized at periastron, when the ultraviolet photon field is particularly dense and the intrinsic cut-off energy is the highest. 

In the case of an isotropic radiation field the optical depth could reach $\sim 10$ at periastron. As the soft photon distribution is highly anisotropic, the absorption cross section will decrease and peak up to $\epsilon_1\epsilon_2>10\ {\rm MeV}^2$ for $\gamma$-ray photons going away from the stars towards the observer \citep{2018MNRAS.474.1436V}. This is consistent with the H.E.S.S. observations \citep{2017arXiv170801033L} showing a cut-off at an energy higher than expected for an isotropic distribution of ultraviolet photons. The $\gamma$-ray spectrum cut-off energy and optical depth are expected to vary along the orbit (and viewing angle), showing a variability pattern indicative of the geometry and magnetic field configuration. 

CTA will follow these variations with a sensitivity which is orders of magnitude better than \flat, and will provide unique additional constrains on the model parameters. Provided that the loosely constrained optical depth is in a reasonable range, the $\pi^0$-decay continuum could be detected up to 100 TeV with the small CTA telescopes. Fig. {\ref{fig:sed}} shows that CTA could independently detect the cutoff of the $\pi^0$-decay and the strength of the $\gamma$-$\gamma$ absorption at apastron and periastron \citep[see][]{2018arXiv181004168W}. CTA could follow the variability of these parameters along the orbit with a resolution of a few days to establish the nature of the high energy component and constrain the geometry of the shock at the core of the system. 

\subsection{Prospective CWBs \label{CWB2}}

To date $\eta$ Car is the only CWB which has been detected as a high and a very high energy $\gamma$-ray source. Based on the results of theoretical modelling \citet{2013A&A...555A.102W}  highlighted a sample of seven other CWBs with WR-companions, -- namely WR 11, WR 70, WR 125, WR 137, WR 140, WR 146, and WR 147 -- as the most favourable candidate high energy CWB sources. 
A \flat analysis of these CWBs, using almost 7 years of data, is presented in \citet{2016MNRAS.457L..99P}. As a result, three sources (WR 11, WR 125, WR 147) have been detected with a significance which exceeds a test statistic of TS=25, but only WR11 (part of the \gamvel binary system) does not suffer from background contamination. WR 11 (\gamvel) is also the only WR star which has a counterpart in the FL8Y Source List (FL8Y J0809.4-4714, 6.8 $\sigma$ detection in the 100\,MeV-1\,TeV band), lying within the $3 \sigma$ error ellipse. 

All other WR stars, including WR 125 and WR 147, do not have FL8Y counterparts. Therefore, of the proposed list of seven candidate sources, we will only investigate \gamvel (which consists of WR 11 and an O7.5 star; see Table \ref{parameters-cwb}) 
as a prospective TeV CWB.

We will also consider one further CWB candidate, the binary system HD 93129A, which consists of two  O-type stars (see Table \ref{parameters-cwb}). We have found that the position of HD 93129A is well inside the 2$\sigma$ error ellipse of a FL8Y source, FL8Y J1043.6-5930 (5$\sigma$ detection), which is a counterpart to 3FGL J1043.6-5930.

\subsubsection{$\gamma^2$ Velorum}

\begin{figure*}[h]
\includegraphics[width=0.94\columnwidth]{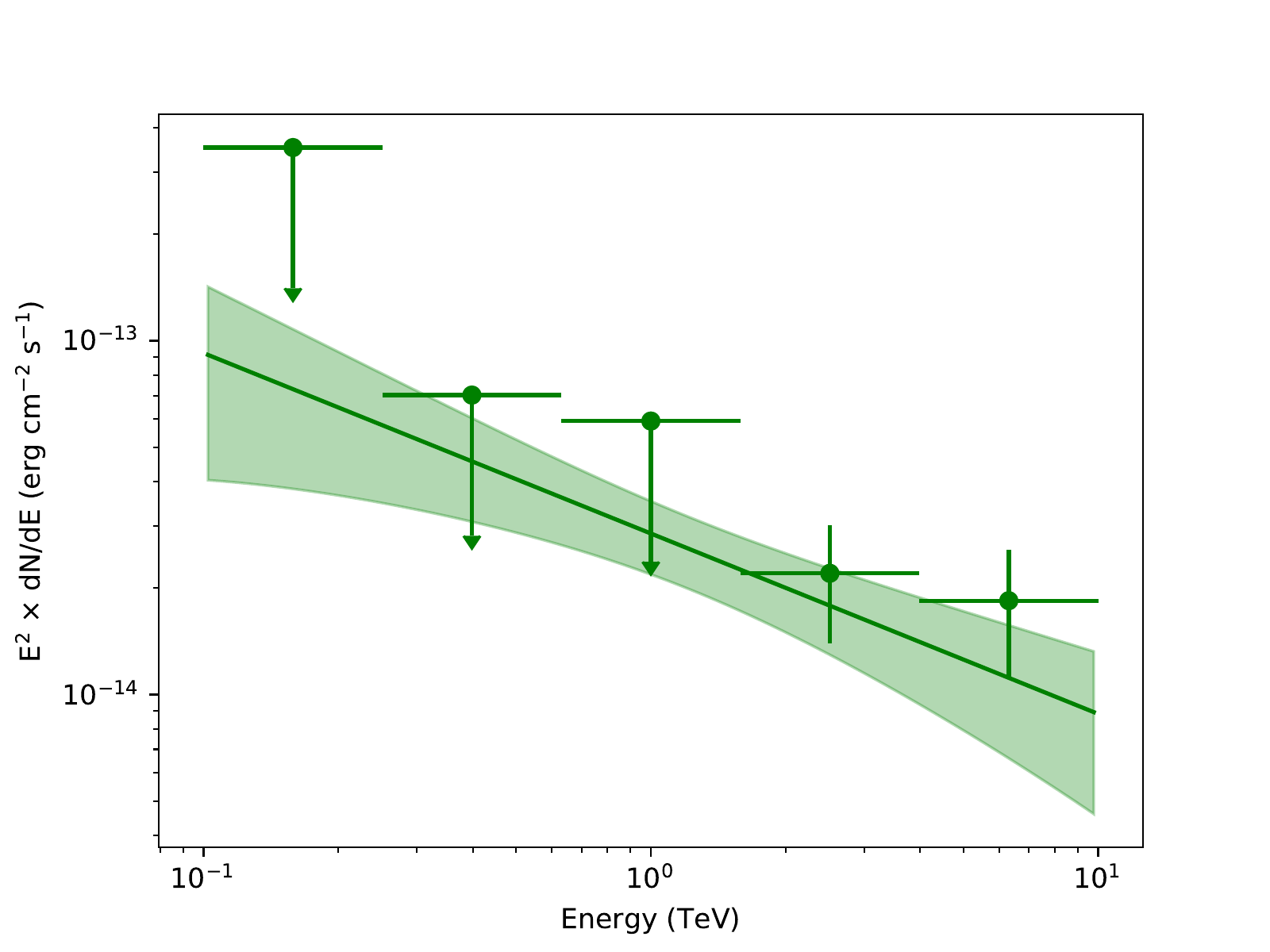}
\includegraphics[width=0.94\columnwidth]{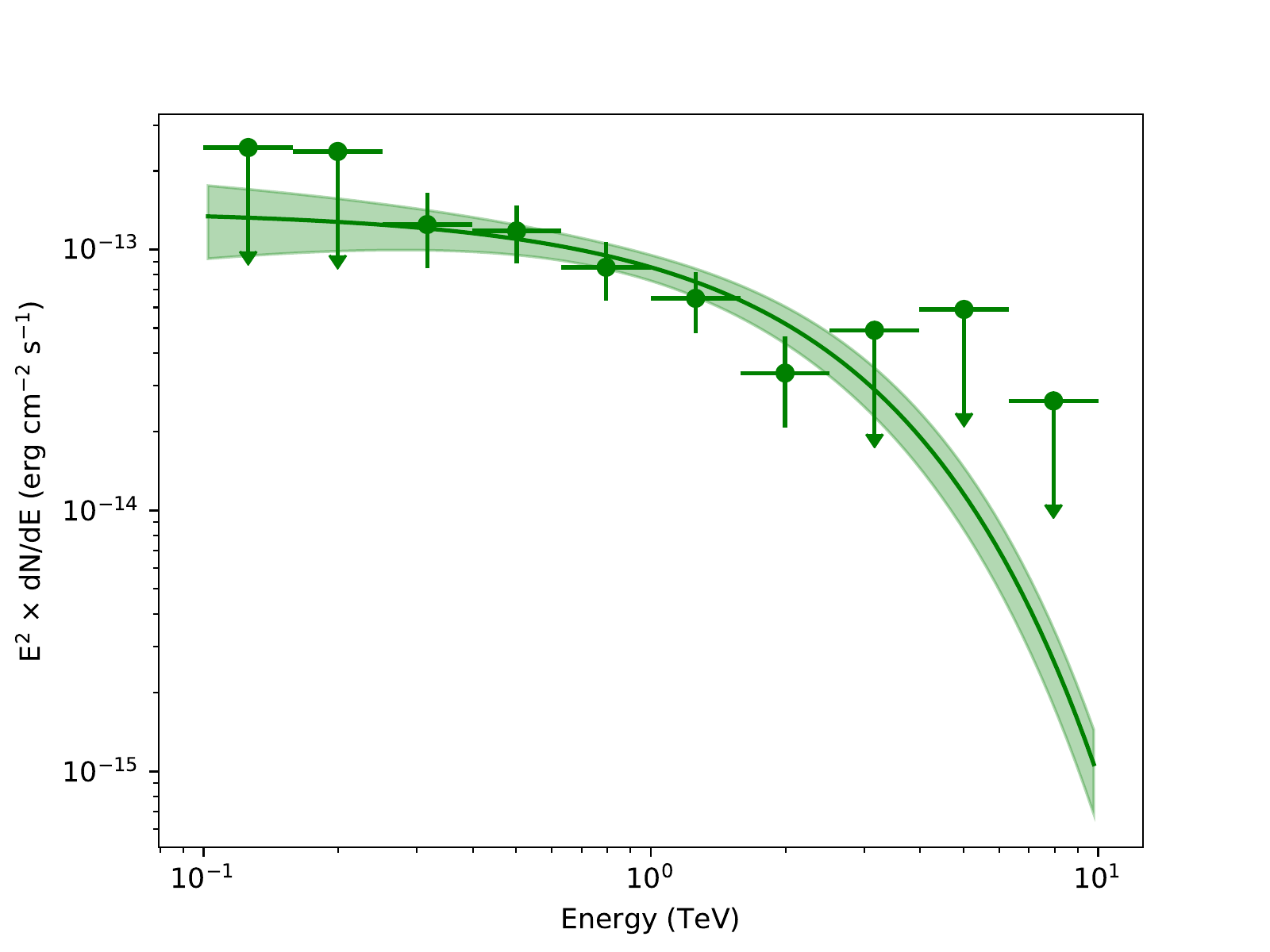}
\caption{Simulated spectra of \gamvel for extrapolated FL8Y power-law input spectrum (\emph{left}) and for additional hard component (power-law with $\Gamma=2.0 $ and with $E_{\rm cut}$=2 TeV cut-off, \emph{right}). 
A 50\,h observation was used for the simulation.
}
\label{fig:gam2vel}
\end{figure*} 

The system \gamvel is one of the most promising CWB candidates for 
detection at HE and VHE gamma rays. Besides the powerful winds given off by both components 
(a WR and a O-type star)
and a small binary separation (see Table \ref{parameters-cwb}), its main advantage is its close proximity: at only 340\,pc, WR11 is the closest WR star to the Earth. Therefore, this binary is very well studied at different energy bands (see e.g.\ references in the Catalogue of Particle-Accelerating Colliding-Wind Binaries, \citealt{2017A&A...600A..47D}).

Since the PSF of a \flat source is on the order of $\sim 1^\circ$, the analysis of high spatial resolution radio data are very important for confirming the high energy $\gamma$-ray association since radio observations can spatially separate additional sources like nearby AGNs.  
Signatures of strongly attenuated non-thermal radio emission from this CWB have been revealed \citep{1999ApJ...518..890C} confirming (with caution) the status of \gamvel as particle-accelerating colliding-wind binary. Note, however, that in the region of \flat enhancement there is another potential high energy $\gamma$-ray source: an extended (45") radio source MOST0808-471 with a non-thermal radio-spectrum and parameters typical of a  FRI or FRII radio galaxy \citep{1999ApJ...518..890C}. The absence of variability in the GeV emission does not allow a firm conclusion to be made on the nature of the observed GeV emission. X-ray (ASCA) and $\gamma$-ray (INTEGRAL) observations place only upper limits on a possible non-thermal component from \gamvel \citep{2004ESASP.552..409T}.

The HE $\gamma$-ray spectrum of \gamvel, using 7 years of \flat data (in the 0.1-100 GeV energy range) was analysed in \citet{2016MNRAS.457L..99P}. It was shown that
fits with a power-law model ($\Gamma=2.16 \pm 0.2$; TS=37.3), a log-parabola model (TS=41.5), and a broken power-law model (TS=44.3) do not well represent the observed hardening of the spectrum at energies $E>$ 10 GeV (this hardening is similar to the spectrum of $\eta$ Car during periastron).
At a distance of $d$=340 pc, the high energy flux of $F$(0.1-100 GeV)  = (2.7$\pm$0.5)$\times$ 10$^{-12}$ erg cm$^{-2}$ s$^{-1}$ corresponds to a luminosity of $L$=(3.7$\pm$0.7)$\times$ 10$^{31}$ erg s$^{-1}$. This is only a small fraction ($\sim$ 10$^{-4}$) of the wind kinetic power dissipated in the colliding wind zone  \citep{2016MNRAS.457L..99P,2017ApJ...847...40R}.
 
The first 3 dimensional MHD simulations of the colliding wind region in \gamvel, which took into account the generation of $\gamma$-ray emission via diffusive shock acceleration of protons and nuclei, and subsequent pion decay,  is presented in \citet{2017ApJ...847...40R}. The \flat data \citep{2016MNRAS.457L..99P} can only be reproduced using a high WR mass-loss rate ($\dot M_{WR}$ = 3$\times$10$^{-5}$ M$_{\odot}$ yr$^{-1}$), but the observed hardening of the $\gamma$-ray spectrum at 10-100 GeV is not reproduced since the simulated spectrum has a cut-off at 100 GeV due to the accelerated protons reaching a maximum energy of $\sim1$\,TeV.

\subsubsection{Prospects for CTA observations}
 
To estimate CTA's capability of detecting \gamvel at TeV energies we extrapolate the \flat spectrum to the  0.1--10 TeV range. In our analysis we assumed  $F$($E$=1 TeV) = 1.77$\times$ 10$^{-20}$ ph cm$^{-2}$ s$^{-1}$ MeV$^{-1}$ and a photon index $\Gamma =2.51$. A simulated spectrum  of a 50\,h observation is shown in Fig. \ref{fig:gam2vel} (left panel).
Even if the \flat power-law spectrum really does extends up to 10 TeV, due to the low flux level the source can only be detected, with a low significance (10<TS<25), in the range  1 -- 10 TeV (where the CTA-sensitivity is maximal).

In order to investigate CTA's ability to detect a hard VHE component from possible pp-interactions in the colliding wind region, we extrapolate the hardening of the \flat spectrum, revealed in \citet{2016MNRAS.457L..99P}, into the TeV range. We  simulate this spectral component assuming a power law spectrum with $F$($E$=1 TeV) = 8.8 $\times$ 10$^{-20}$ ph cm$^{-2}$ s$^{-1}$ MeV$^{-1}$, a photon index $\Gamma =2.0$ and (inspired by $\eta$ Car observations) an exponential cut-off with $E_{cut}$ = 2 TeV (see right panel of Fig. \ref{fig:gam2vel}). This is the most promising case, with a detection about 11$\sigma$ (TS=130) with a 50\,h observation.

\subsubsection{HD 93129A \label{HD}}
\begin{figure}
\includegraphics[width = \columnwidth]{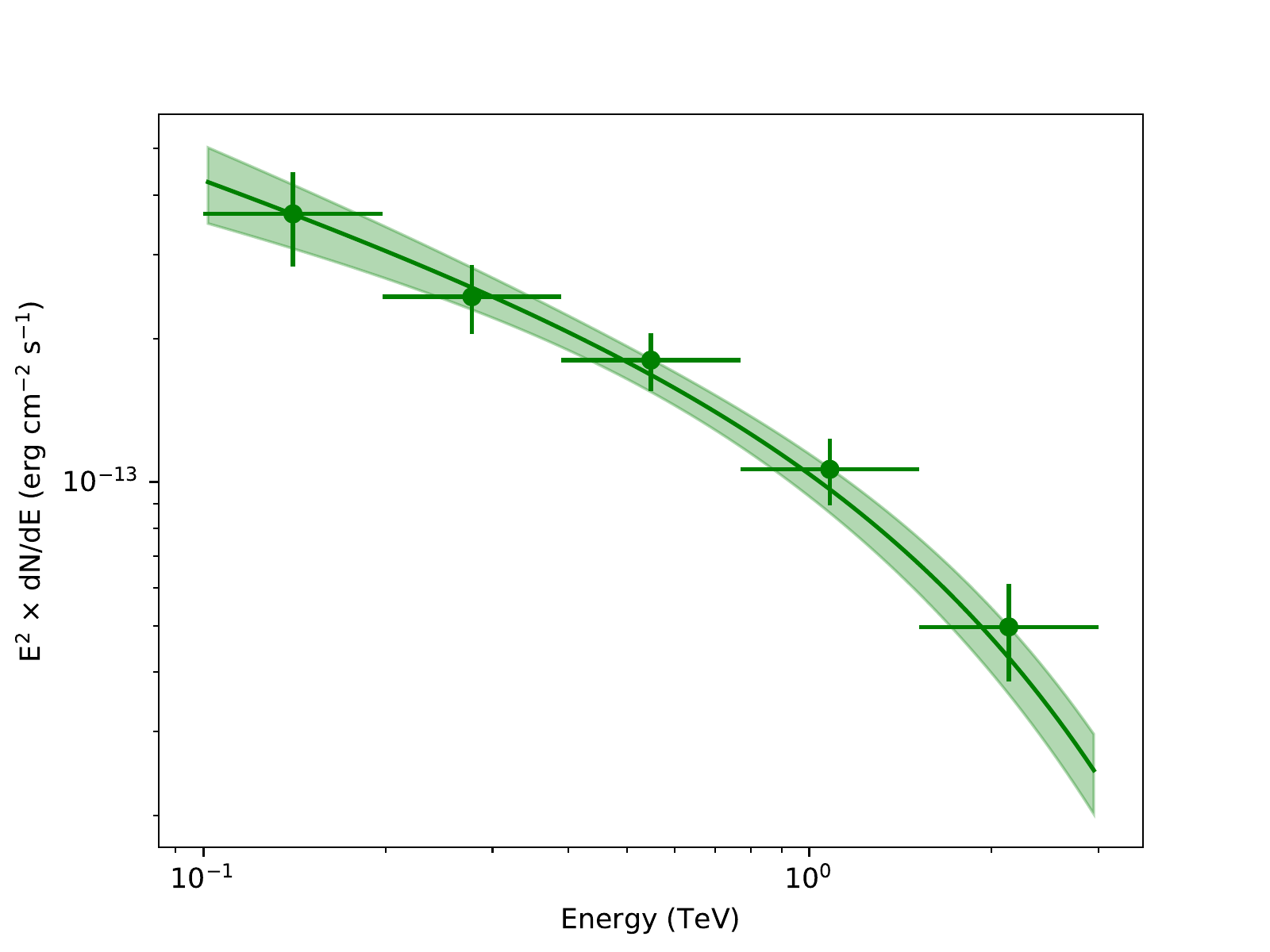}
\caption{Simulated spectrum of HD 93129A for an extrapolated FL8Y power-law with $E_{\rm cut}$=2 TeV cut-off input spectrum.
A 50\,h observation was used for the simulation.
}\label{fig:HD_pl}
\end{figure}

Another potentially detectable TeV CWB system is HD 93129A, one of the most massive binaries in the Galaxy (the masses of the components are approximately 100 and 70 solar masses). Preliminary estimates of the binary period (more than 50 yr) and the orbital inclination angle ($i\sim 15 \degr$) predicted a periastron passage in 2020 \citep{2016A&A...591A.139D,2011ApJS..194....5G}. Improved calculations in \citet{2017MNRAS.464.3561M} suggest a long-period orbit with e > 0.92 and with the periastron passage occurring in 2017/2018 (Table \ref{table:parameters-cwb}).

Non-thermal radio emission from the colliding wind region in HD 93129A was detected by VLBI \citep{2016A&A...591A.139D}, while X-ray emission from the embedded wind shocks around HD 93129A (the primary component) and the colliding wind region, was detected during the Chandra Carina Complex survey \citep{2011ApJS..194....5G}.

In \citet{2017AIPC.1792d0027D} a detailed investigation of the high-energy emission from the CWB HD 93129A is presented and it is shown that for some sets of allowed parameters the expected VHE fluxes can be detected with CTA. Note that the position of HD 93129A is well inside the 2$\sigma$ error ellipse of FL8Y J1043.6-5930.

In order to investigate the prospects of CTA observations, we extrapolated the \flat spectrum
[$F$($E$) =(3.7 $\pm$ 0.7)$\times$ 10$^{-14}$ $(E/E_p)^{-\Gamma}$ ph cm$^{-2}$ s$^{-1}$ MeV$^{-1}$, $E_p$ = 5 GeV , $\Gamma$ = 2.4 $\pm$ 0.1] to the TeV band and assumed an exponential cut-off at  $E_{\rm cut}$ = 2 TeV.  A simulation of a 50 hour observation results in a detection with a total test statistic of TS=190 in the 100 GeV - 3 TeV band (Fig \ref{fig:HD_pl}).

\section{Discussion and Conclusions}
\label{sec::summary}

In this paper we presented an overview of the current observations of $\gamma$-ray binaries and illustrated the CTA capability to study the spectral and timing properties of these sources. The full list of the considered sources with the basic parameters of the hosting systems are summarized in Table~\ref{parameters} and Table~\ref{table:parameters-cwb}.

Despite dedicated observations with all modern VHE facilities (H.E.S.S., MAGIC, VERITAS, HAWC) the statistics of the VHE data is still poor.  At best, $\gamma$-ray binaries can be only marginally detected at 
VHE in a few hours, 
whereas for some of them variability on time scales as short as $\sim30$~minutes is observed in the X-ray and HE  bands \citep[e.g.][]{2009MNRAS.397.2123C, 2018ApJ...863...27J}. 
The measurements of the basic spectral parameters (e.g. spectral slope) in the TeV band require even longer exposure times. The studies of more complex parameters, such as spectral curvature, orbital variations of the cut-off energy, orbit-to-orbit variability and short time scale variability in most cases are beyond the current sensitivity limits. This complicates a comparison of TeV and lower-energies (radio to GeV) data where all listed above variabilites are known to be present. 
The large uncertainty in the study of simultaneous variability prevents a firm confirmation of whether or not the same relativistic particle population is responsible for the broadband emission. 
See also~\citet{paredes19} for a broader list of open questions.

The unprecedented sensitivity of CTA will, in most cases,  enable the spectral evolution of such binaries as \psrb, \ls, \lsi and HESS J0632+057  to be studied on time scales as short as 30 min, comparable to the variability observed in X-rays. 
This will 
address the statistic-related issues of the current-generation instruments described above and provide input data for theories describing details of particle acceleration. For example the presence of clumps in a stellar wind is thought to lead to the modification of the wind-wind collision shock structure and affect the efficiency of particle acceleration in relativistic hydrodynamical simulations~\citep[see e.g.][]{paredes-fortuny15,dubus15}.  
The CTA observations of the short-time scale variability in $\gamma$-ray binaries may allow the the properties of the clumps to be assessed.
For example their  scale-distrubution can be reconstructed using the approach similar to the one applied by \citet{Chernyakova2017} to X-ray data.

Studies of the orbital variability of the high-energy spectral cut-off can  
identify the 
environmental parameters which lead to the most efficient particle acceleration. Current observations are unable to constrain the maximal energy of relativistic particles for most of the known binaries. Only in the case of \ls is the cut-off energy $E_{cut}=6.6$ TeV measured close to the inferior conjunction \citep{Mariaud15}. Under the assumption  that the $\gamma$-ray emitter is a jet-like structure \cite{LS5039_2008_Khan} built a model reproducing the spectral states of the system near the inferior and exterior conjunctions. However, to test this  model  and constrain the physical properties of the source observations capable of providing detailed energy spectra for narrow orbital phase intervals ($\Delta \varphi < 0.1$) are needed. As it is shown by the simulations presented in Fig.~\ref{fig:binaries} and in Section 3, CTA will be easily able to  
undertake such observations.

Below we briefly summarize what CTA could do for $\gamma$-ray binaries   and outline the  questions that can be addressed.

\underline{\psrb}: CTA observation around the periastron passage will allow the spectral variability to be studied on a 30 min timescale at a level of better than 5\%. This will finally answer the question of whether the TeV light curve also has two peaks around the periastron, similar to radio and X-rays, and give us a chance to study spectral evolution due to gamma-gamma absorption.
Accompanied with simultaneous multi wavelength observations CTA data will allow us to test numerous models of particle acceleration in the system, the composition of the pulsar wind and the details of the interaction of the pulsar wind with the disk of the Be star leading to the still unexplained GeV flare. 

\underline{\lsi, \ls and \fgl}: it will be possible to perform spectral studies with an accuracy of better than 10\% on a few hours time scale, which is less than one percent of the orbital period of these systems. CTA observations will also study the short time scale variability  down to 30 min, which will allow for a direct multi wavelength comparison. Detailed reconstruction of the broad band spectral variability on short and orbital time scales will allow us to test existing models and reconstruct the details of the geometry and physical conditions of the emission region. 

\underline{HESS J0632+057}: CTA observations will allow the high energy slope and energy break  to be accurately monitored  along the orbit. 50 hours of observation will allow the flux and slope of the source to be reconstructed with a better than 3\% accuracy, and the position of a spectral break with a 10\% accuracy. This will enable us to disentangle 
the current theoretical models, to understand if the compact object in the system undergoes accretion close to periastron ("flip-flop" model), or if the observed two peak orbital light curve is due to the inclination of the Be star disk to the orbital plane.

\underline{HESS J1832-093}: for less studied binary systems, like this one, CTA observations will be crucial for understanding their nature. Indeed, CTA observations of HESS J1832-093, with its improved angular resolution will place further constraints on the extension of the TeV source. This information, accompanied with the detailed measurements of the spectral variability of the source on different time scales, will help to ascertain the nature of the system and find out whether it is indeed a \grb. 

\underline{LMC P3}: CTA sensitivity will be high enough to study for the first time the nightly averaged spectral evolution of the source with an accuracy better than 10\% along the orbit. This will put severe constraints on the geometry of the system and on the details of the physical processes governing the observed emission, and will contribute to the development of a consistent model for the multi wavelength emission.

\underline{\etacar and other CWBs:}  we demonstrated that CTA will be able to greatly advance the study of high energy emission for CWBs. At the time of writing TeV emission was detected from only one CWB, namely $\eta$ Car. With CTA it will be possible to probe the spectrum and variability of $\eta$ Car above 30 GeV along the orbit with a resolution of a few days. This will allow the nature of the high energy component to be established and constrain the geometry of the shock at the core of the system. For other CWBs candidates, with \gamvel and HD 93129A the most prominent, CTA sensitivity will be high enough to probe the presence of TeV emission.

\begin{acknowledgements}

          We gratefully acknowledge financial support from the agencies and organizations 
          listed here: \\
          
\href{http://www.cta-observatory.org/consortium_acknowledgments}{http://www.cta-observatory.org/consortium\_acknowledgments}.

This research made use of {\tt ctools}, a community-developed analysis package for Imaging Air Cherenkov Telescope data. 
ctools is based on {\tt GammaLib}, a community-developed toolbox for the high-level analysis of astronomical $\gamma$-ray data. \\ 
This research has made use of the CTA instrument response functions provided by the CTA Consortium and Observatory, 
see \href{https://www.cta-observatory.org/science/cta-performance/}{https://www.cta-observatory.org/science/cta-performance/} 
(version prod3b-v1) for more details. \\ 
The authors acknowledge support by the state of Baden-W\"urttemberg through bwHPC. This work was supported by the Carl-Zeiss Stiftung through the grant ``Hochsensitive Nachweistechnik zur Erforschung des unsichtbaren Universums'' to the Kepler Center f{\"u}r Astro- und Teilchenphysik at the University of T{\"u}bingen.\\
PR, GP, FC and SV acknowledge contribution from the grant INAF CTA--SKA, 
``Towards the SKA and CTA era: discovery, localisation, and physics of transient sources'' (PI M.\ Giroletti). \\
LS acknowledges financial support of the ERDF under the Spanish MINECO (FPA2015-68378-P and FPA2017-82729-C6-3-R). \\
JMP and MR acknowledge support by the Spanish Ministerio de Economía, Industria y Competitividad (MINEICO/FEDER, UE) under grants FPA2015-69210-C6-2-R, AYA2016-76012-C3-1- P, MDM-2014-0369 of ICCUB (Unidad de Excelencia ‘María de Maeztu’) and the Catalan DEC grant 2017 SGR 643.\\
PLE and JM also acknowledge support by MINECO/FEDER, UE under grant AYA2016-76012-C3-3-P
and by the Consejer\'{\i}a de Econom\'{\i}a, Innovaci\'on, Ciencia y Empleo of Junta de Andaluc\'{\i}a under research group FQM-322.\\
MC acknowledge SFI/HEA Irish Centre for High-End Computing (ICHEC) for the provision of computational facilities and support.

\end{acknowledgements}



\end{document}